\documentclass[11pt,letterpaper]{article}
\usepackage{systeme} %para usar sistemas de ecuaciones
\usepackage[utf8]{inputenc}
\usepackage{multirow} %para las tablas
\usepackage{booktabs} %para la estética de tablas
\usepackage[english]{babel}
\usepackage{amsmath}
\usepackage{amsfonts}
\usepackage{amssymb}
\usepackage{lmodern}
\usepackage{graphicx}
\usepackage[small,bf]{caption} %Para caption con letra pequeña y Figura en negrita
\usepackage[margin=2.5cm]{geometry}
\usepackage{framed} %Para cuadro de texto
\usepackage{color} %Para cuadro de texto
\usepackage{wrapfig}\definecolor{shadecolor}{RGB}{220,220,220} %Para color cuadro de texto, codigo 220,220,220 es gris
\usepackage{float} %Pone una imagen exctamemte donde quiero
\usepackage{array} %para poner harto texto en tablas
\usepackage{caption} %Para que algunas tablas no tengan numeración
\usepackage{microtype}
\usepackage{titlesec}
\usepackage{tcolorbox} %Para recuadro con color en las ecuaciones
\titleformat{\section}
  {\large\bfseries}
  {\thesection}{1em}{}
\titleformat{\subsection}
  {\normalsize\bfseries}
  {\thesubsection}{1em}{}
\titlespacing*{\section}{0pt}{16pt}{6pt}
\titlespacing*{\subsection}{0pt}{12pt}{4pt}

\author{Jorge Pinochet}
\title{\textbf{From Geometry to Gravity: A Friendly Introduction \\ to the Schwarzschild Metric}}

\begin{document}

\author{Jorge Pinochet$^{*}$\\ \\
 \small{$^{*}$\textit{Facultad de Ciencias Básicas, Departamento de Física. }}\\
  \small{\textit{Centro de desarrollo de Investigación CEDI-UMCE,}}\\
 \small{\textit{Universidad Metropolitana de Ciencias de la Educación,}}\\
 \small{\textit{Av. José Pedro Alessandri 774, Ñuñoa, Santiago, Chile.}}\\
 \small{e-mail: jorge.pinochet@umce.cl}\\}

\date{} %NO PONER FECHA
\maketitle

\begin{center}\rule{0.9\textwidth}{0.1mm} \end{center}
\begin{abstract}
\noindent The first exact solution to the field equations of general relativity was found by the German physicist and astronomer Karl Schwarzschild in 1916. Due to its simplicity and its various astronomical applications, the Schwarzschild solution is the most important, which gives it great educational potential. However, this potential is not usually exploited, since the Schwarzschild solution is generally taught in advanced undergraduate physics courses or in specialized graduate courses. The objective of this article is to offer a friendly introduction to the Schwarzschild solution, and at the same time analyze some concepts of Riemannian geometry and general relativity, employing only simple mathematical tools that do not go beyond the rudiments of differential and integral calculus.\\


\end{abstract}

\maketitle

\section{Introduction}
General relativity (GR) is the theory of gravity proposed by Einstein in 1915 to refine and extend Newton’s law of universal gravitation [1–3]. GR is a geometric theory of gravity that replaces Newtonian attractive forces with the concept of curvature or distortion of space-time [1] \footnote{It is worth mentioning that Einstein’s initial motivation was not to generalize Newtonian gravitation, but rather to extend Special Relativity to non-inertial systems}. This means that in Einstein’s universe the force of gravity does not exist, and what guides the motion of planets, stars, and galaxies is the curvature of space-time [2]. When an object moves freely from one place to another, which implies the presence of gravity, it simply follows the trajectories dictated by the curved geometry. Consequently, unlike what occurs in Newtonian physics, where space and time are absolute and immutable, in GR they are dynamic and flexible entities that respond to the presence of matter.\\

For a long time only a few exact solutions to the equations of GR were known, but in recent decades many others have been found through new computational procedures. Most of the solutions have purely mathematical interest and lack relevant physical or astronomical applications. The first exact solution was found by the German physicist and astronomer Karl Schwarzschild in 1916, shortly after Einstein published his revolutionary theory of gravity [3]. Due to its relative simplicity and its various physical and astronomical applications, the Schwarzschild solution or \textit{Schwarzschild metric} is one of the most important exact solutions of Einstein’s field equations, and therefore it has great educational potential.\\ 

However, this potential is not usually exploited, since the Schwarzschild metric is generally taught in advanced undergraduate physics courses or in specialized graduate courses, and introductory-level students usually have little contact with this topic. The reason is that GR is an extremely technical subject whose study requires advanced mathematical knowledge. However, it is often forgotten that it is not necessary to introduce the entire complex mathematical apparatus of GR in order to understand the Schwarzschild solution and apply it in contexts of physical and astronomical interest. Moreover, in the author’s experience, the understanding of GR in graduate studies is strongly favored when students arriving at that stage are already familiar with the Schwarzschild solution and understand its meaning and its physical and astronomical applications.\\

The objective of this work is to offer a friendly introduction to the Schwarzschild solution, which will lead us to explore some concepts of Riemannian geometry and relativistic physics. To introduce these concepts we will employ only mathematical tools that do not go beyond the rudiments of differential and integral calculus. Although the material presented in this work is not new, its main contribution lies in the way it is organized and articulated for pedagogical purposes. From this perspective, the work can be used, for example, as an educational resource in undergraduate courses in astronomy or modern physics, where the analysis of the Schwarzschild solution can follow the study of the theory of special relativity and Minkowski flat space-time.\\

The article is organized as follows. In Section 2 the concept of metric in Riemannian geometry is introduced. In Section 3 the notions of curvature and geodesic are discussed from an intuitive point of view. The following sections gradually lead to the introduction of the Schwarzschild metric and to some of its physical applications. The article is intended for readers who master the theory of special relativity, the rudiments of differential and integral calculus, and who are familiar with GR.

\section{Riemannian Geometry and the Concept of Metric}

Euclidean geometry is the branch of mathematics that studies the properties of flat spaces. Curved spaces are described by non-Euclidean geometry [4]. As we will see in more detail in Section 3, the essential difference is that only Euclidean geometry satisfies the celebrated \textit{parallel postulate}, which states that two lines that are initially parallel, extended indefinitely, maintain their mutual distance constant.\\ 

The branch of geometry that studies both flat and curved spaces in any dimension is \textit{Riemannian geometry}, which is the mathematical foundation of GR, and contains Euclidean geometry as a particular case of zero curvature. Its fundamental object of study is the \textit{manifold}, which extends the notions of curve (one-dimensional manifold) and surface (two-dimensional manifold) to spaces of $n$ dimensions ($n$-dimensional manifolds) [5]. The fundamental principle of Riemannian geometry is: \textit{In a small neighborhood of a point, non-Euclidean manifolds agree with Euclidean geometry}. In other words, a sufficiently small region of a curved space can be considered flat in first approximation [5]. An important corollary of this principle is that the Pythagorean theorem of Euclidean geometry is locally valid on a non-Euclidean manifold.\\ 

\begin{figure}[h]
  \centering
    \includegraphics[width=0.3\textwidth]{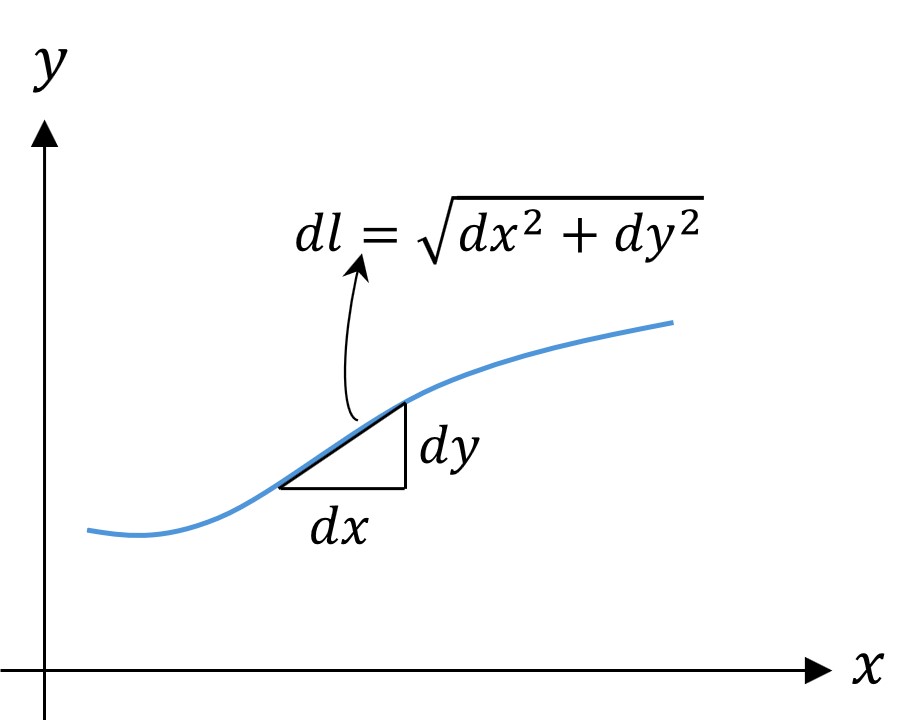}
  \caption{The Pythagorean theorem written in differential form.}
\end{figure}

The fundamental mathematical object of Riemannian geometry is the \textit{metric}, which is a powerful generalization of the Pythagorean theorem. The metric allows the calculation of the infinitesimal distance or \textit{line element} between two points on any manifold. To explore Riemannian geometry more deeply, let us analyze some simple examples that allow us to illustrate the power and usefulness of the concept of metric.\\

\begin{figure}[h]
  \centering
    \includegraphics[width=0.3\textwidth]{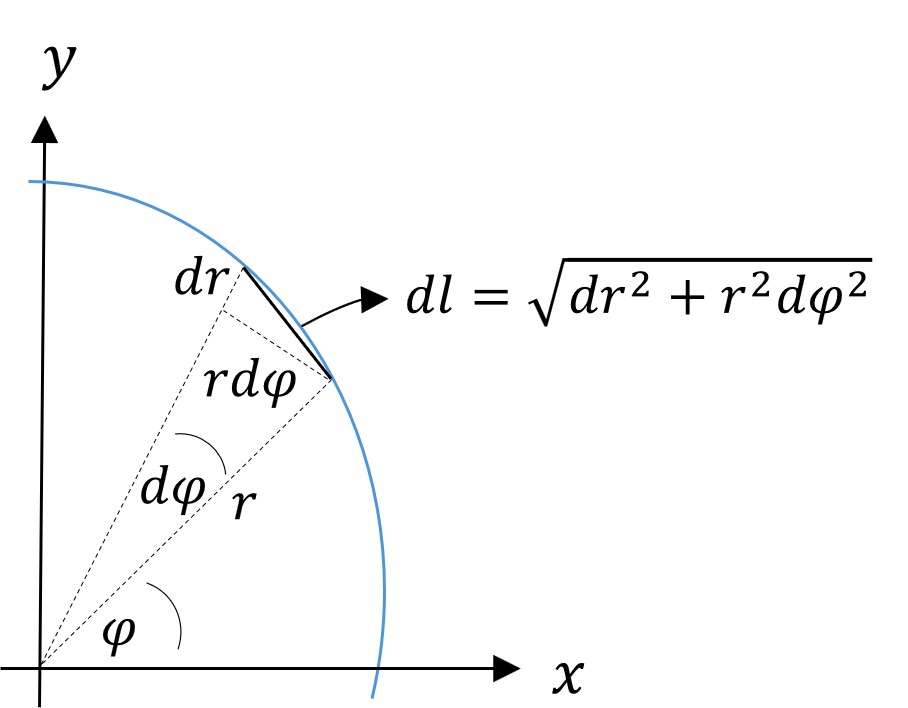}
  \caption{Metric in polar coordinates.}
\end{figure} 

As a first example, consider two points on a two-dimensional Euclidean manifold (flat surface). According to the Pythagorean theorem (Fig. 1), the square of the line element between these points in Cartesian coordinates $(x,y)$ is

\begin{equation} % 1
(dl)^{2}=(dx)^{2}+(dy)^{2}.
\end{equation}

Adopting the convention of removing parentheses yields

\begin{equation} % 2
dl^{2}=dx^{2}+dy^{2}.
\end{equation}

This is the metric of the two-dimensional Euclidean manifold, which we can rewrite as

\begin{equation} % 3
dl^{2}= (1)dxdx + (1)dydy + (0)dxdy + (0)dydx,
\end{equation}

or equivalently,

\begin{equation} %4
dl^{2} = g_{11}dx_{1}dx_{1} + g_{22}dx_{2}dx_{2} + g_{12}dx_{1}dx_{2} + g_{21}dx_{2}dx_{1},
\end{equation}

where $x_{1}=x$, $x_{2}=y$; $g_{11}=g_{22}=1$, $g_{12}=g_{21}= 0$, so that the metric is expressed in the compact form

\begin{equation} %5
dl^{2} = \sum_{i=1}^{2} \sum_{j=1}^{2} g_{ij} dx_{i} dx_{j},
\end{equation}

and the $g_{ij}$ can be represented by a symmetric $2\times2$ matrix:

\begin{equation} %6
\left[g_{ij} \right] = 
\begin{pmatrix}
g_{11} & g_{12}\\
g_{21} & g_{22} 
\end{pmatrix}
= \begin{pmatrix}
1 & 0\\
0 & 1 
\end{pmatrix}.
\end{equation}

As a second example, consider the metric for the same manifold as above, but described in polar coordinates $(r,\varphi)$ (see Fig. 2):

\begin{equation} %7
dl^{2}= dr^{2}+ r^{2}d\varphi ^{2}.
\end{equation}

As the reader can easily verify, using the same ideas that led us to Eqs. (5) and (6), we find that the corresponding metric is also given by Eq. (5), except that now $x_{1}=r,x_{2}=\varphi; g_{11}=1,g_{22}=r^{2}; g_{12}=g_{21}=0$, where 

\begin{equation} %8
\left[g_{ij} \right] = 
\begin{pmatrix}
1 & 0\\
0 & r^{2} 
\end{pmatrix}.
\end{equation}

As a third example, consider a three-dimensional Euclidean manifold, that is, a space equivalent to the one we inhabit, and write the metric in Cartesian coordinates $(x,y,z)$:

\begin{equation} %9
dl^{2} = dx^{2}+dy^{2}+dz^{2}.
\end{equation}

Once again the reader can verify that the metric is described by an expression analogous to Eq. (5), where only the range of variation of the subscripts $i$ and $j$ changes:

\begin{equation} %10
dl^{2} = \sum_{i=1}^{3} \sum_{j=1}^{3} g_{ij} dx_{i} dx_{j}.
\end{equation}

In this case $x_{1}=x,x_{2}=y,x_{3}=z; g_{11}=g_{22}=g_{33}=1; g_{ij}=0$ if $i\neq j$, where

\begin{equation} %11
\left[g_{ij} \right] = 
\begin{pmatrix}
1 & 0 & 0\\
0 & 1 & 0\\
0 & 0 & 1
\end{pmatrix}.
\end{equation}

Let us consider one final example, which will be useful when we discuss four-dimensional relativistic geometry. If we generalize Eq. (9) to a four-dimensional Euclidean space described in Cartesian coordinates $(x,y,z,\omega)$, the metric can be written as

\begin{equation} %12
dl^{2} = dx^{2}+dy^{2}+dz^{2}+d\omega^{2},
\end{equation}

and the reader can verify that

\begin{equation} %13
dl^{2} = \sum_{i=1}^{4} \sum_{j=1}^{4} g_{ij} dx_{i} dx_{j},
\end{equation}

where $x_{1}=x,x_{2}=y,x_{3}=z,x_{4}=\omega; g_{ij}=1$ if $i=j$; $g_{ij}=0$ if $i\neq j$, where 

\begin{equation} %14
\left[g_{ij} \right] = 
\begin{pmatrix}
1 & 0 & 0 & 0\\
0 & 1 & 0 & 0\\
0 & 0 & 1 & 0\\
0 & 0 & 0 & 1
\end{pmatrix}.
\end{equation}

We could continue introducing other examples of metrics, both for Euclidean and non-Euclidean manifolds, and we would discover that it is always possible to write them in the concise form given by Eq. (5), but modifying the range of variation of the subscripts according to the dimension of the manifold. Einstein realized that it was possible to be even more concise, since whenever a formula indicates summation with respect to two subscripts, such as $i$ and $j$ in Eqs. (5), (10), and (13), they necessarily appear twice. This leads to the so-called \textit{Einstein summation convention}: \textit{When a subscript appears repeated in a product, it is understood that one must sum over it, and therefore it is possible to omit the symbol} $\Sigma$ [6]. The summation convention allows us to rewrite Eqs. (5), (10), and (13) as

\begin{equation} %15
dl^{2}=g_{ij}dx_{i}dx_{j}.
\end{equation}

In GR, this expression takes the same form, except that the line element is written as $ds^{2}$. Any metric, regardless of the manifold it describes, can be written in this compact form. Technically, Eq. (15) is said to be \textit{covariant}, that is, it does not change its form under arbitrary coordinate transformations (Appendix A provides a clarifying example). Thus, as we anticipated, Eq. (15) is a powerful generalization of the Pythagorean theorem, where the $g_{ij}$ are functions of the coordinates. If we consider an arbitrary coordinate system $(x_{1},x_{2},...x_{n})$, the $g_{ij}$ can always be represented by a square $n\times n$ matrix called the \textit{metric tensor}, which has $n^{2}$ components, where $n$ is the dimension of the manifold:

\begin{equation} %16
\left[g_{ij} \right]  = 
\begin{pmatrix}
g_{11} & g_{12} & \cdots & g_{1n} \\
g_{21} & g_{22} & \cdots & g_{2n} \\
\vdots  & \vdots  & \ddots & \vdots  \\
g_{n1} & g_{n2} & \cdots & g_{nn} 
\end{pmatrix}.
\end{equation}

This matrix is symmetric, that is, the components below the main diagonal are equal to those above it $(g_{ij}=g_{ji})$. This implies that the number $N$ of independent components is less than $n^{2}$, and is calculated as

\begin{equation} %17
N=1+2+... n=\frac{1}{2}n(n+1).
\end{equation}

Although we will not go deeper into the concept of tensor, since it is a highly technical topic, it is worth noting that a very important characteristic of tensors is that they allow physical equations to be written in covariant form as tensor equations, so that they do not change their form under general coordinate transformations. Another characteristic that becomes evident from the previous considerations is that tensors are a powerful mathematical tool for condensing and packaging information.\\

One of the most important characteristics of the metric tensor is that, from it, all the properties of a manifold can be calculated by extracting the information stored in the $g_{ij}$ [6]. That is, the metric tensor contains all the information about any given manifold. For example, we can calculate areas, volumes, angles, distances between points, curvature, etc. In particular, since a given manifold can be expressed in different coordinate systems, by using the metric tensor and imposing the condition $dl=dl^{'}$, it is possible to perform a transformation that takes us from the metric expressed in coordinates $(x_{1},x_{2},...x_{n})$ to the metric expressed in other coordinates $(x_{1}^{'},x_{2}^{'},...x_{n}^{'})$ (Appendix A).\\

The metric encodes the information of a given manifold according to the direction in which the infinitesimal displacement $dl$ points. This information changes with orientation in a curved manifold, but it is constant in a flat manifold. When the curvature is zero at every point of the manifold, \textit{it is always possible to find a coordinate system where $g_{ij}=1$ if $i=j$, and $g_{ij}=0$ if $i \neq j$, so that the metric tensor takes the simple diagonal form}

\begin{equation} %18
\left[g_{ij} \right]  = 
\begin{pmatrix}
1 & 0 & \cdots & 0 \\
0 & 1 & \cdots & 0 \\
\vdots  & \vdots  & \ddots & \vdots  \\
0 & 0 & \cdots & 1 
\end{pmatrix}.
\end{equation}

Three particular cases of this expression are Eqs. (6), (11), and (14). Finally, it is important to keep in mind that, although for simplicity we have only considered examples of diagonal metric tensors, where only the components on the main diagonal of the matrix are different from zero, non-diagonal tensors also exist, which in general are associated with coordinate systems whose axes are not mutually orthogonal.

\section{Curvature and Geodesic: An Intuitive View}

The geometric properties of a manifold are \textit{invariant}, that is, they are independent of the coordinate system used to describe them. \textit{Curvature} and \textit{geodesics} are two central properties of Riemannian geometry that are invariant [4]. Like all the properties of a manifold, curvature and geodesics are calculated using the metric. Since these calculations are usually complex, we will analyze them intuitively below, focusing attention on two-dimensional manifolds of constant curvature, such as the surface of a sheet of paper or of a sphere. However, everything we say about these surfaces is valid for manifolds of higher dimensions. \\

\begin{figure}[h]
  \centering
    \includegraphics[width=0.25\textwidth]{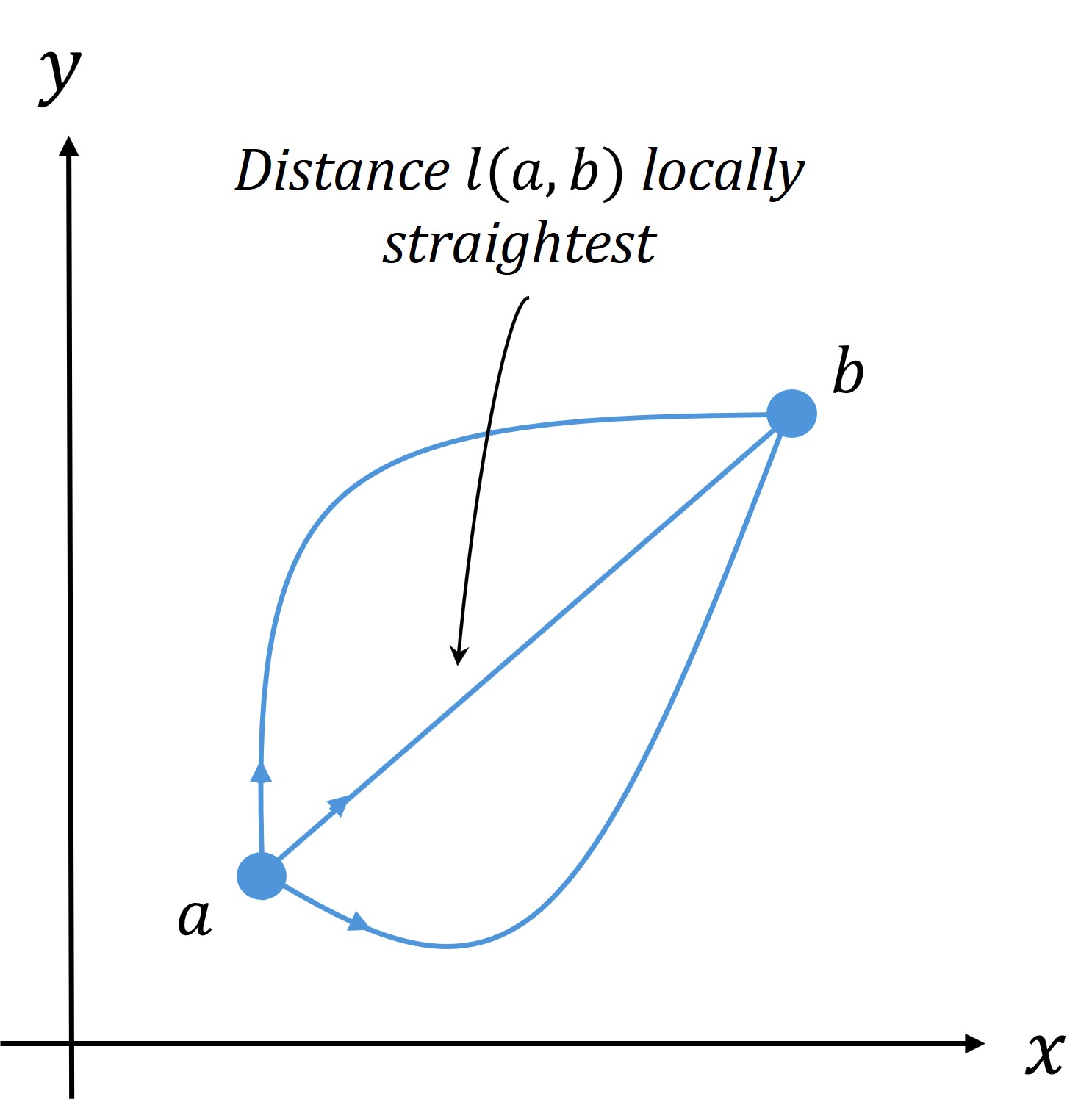}
  \caption{The geodesic $l(a,b)$ is the locally straightest path between two nearby points $a$ and $b$.}
\end{figure} 

Let us begin with the notion of geodesic, which is defined as the locally straightest path between two points on a given manifold, and which is contained within that manifold. In other words, of all the paths that connect two infinitesimally close points on the manifold, the geodesic is the one that most closely approximates a straight line \footnote{In Riemannian geometry, geodesics locally extremize length and, in sufficiently small regions, can minimize distance. In general relativity the situation is somewhat different: timelike geodesics extremize proper time and typically maximize it locally, whereas null geodesics require an affine parameter.} [2]. In a flat manifold, geodesics are straight lines, but in a curved manifold such as the surface of a sphere, geodesics will be curved trajectories (great circles on the surface of the sphere). Thus, the geodesic generalizes the notion of a straight line in the flat geometry of Euclid. \\

\begin{figure}[h]
  \centering
    \includegraphics[width=0.6\textwidth]{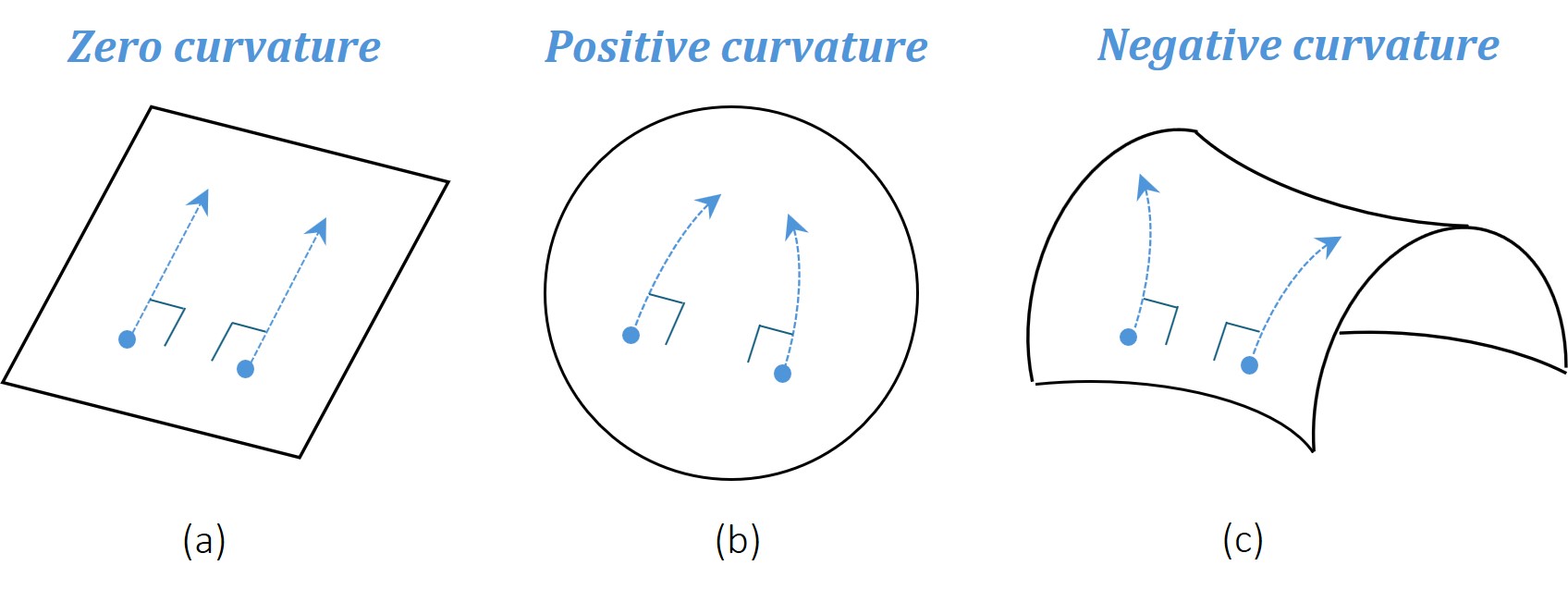}
  \caption{Examples of zero curvature (sheet of paper), positive curvature (sphere), and negative curvature (saddle).}
\end{figure} 

The general procedure for calculating geodesics consists of taking two fixed points $a$ and $b$ on a given manifold, and then determining which of all the paths connecting $a$ and $b$ is locally the straightest (Fig. 3), where the length of each path is calculated from the line element given by Eq. (15):

\begin{equation} %19
\int_{a}^{b} dl = l(a,b) = \int_{a}^{b} \sqrt{g_{ij} dx_{i} dx_{j}}=\int_{a}^{b}\sqrt{g_{ij}\frac{dx_{i}}{dt}\frac{dx_{j}}{dt}}dt,
\end{equation}

where $t$ is a parameter that, from a physical point of view, can be interpreted as the time it takes a test particle\footnote{When we speak of a test particle we refer to a point-like object whose mass is very small, so that it does not significantly modify its surroundings, and its gravity can be neglected.} to move between $a$ and $b$. Intuitively, this equality corresponds to the sum or integration of all the infinitesimal line elements along the trajectory. Thus, the geodesic is the sum of a large number of small straight or almost straight segments. As an example, in the case of a two-dimensional Euclidean manifold, the distance between $a$ and $b$ is

\begin{equation} %20
\int_{a}^{b} dl = l(a,b) = \int_{a}^{b} \sqrt{dx^{2} + dy^{2}}=\int_{a}^{b}\sqrt{\left(\dfrac{dx}{dt} \right)^{2} +\left( \frac{dy}{dt}\right)^{2}}dt.
\end{equation}

Let us now analyze the concept of curvature, which, for simplicity, we will restrict to the case of two-dimensional surfaces of constant curvature (in higher dimensions curvature is not usually described by a single number, but rather by the Riemann tensor). As shown in Fig. 4, a manifold can have only three types of curvature: \textit{zero}, \textit{positive}, and \textit{negative} [7]. Although the surfaces shown in Fig. 4 have constant curvature, it is important to keep in mind that, in most situations, curvature varies from point to point, and therefore it is a local property. Nevertheless, Fig. 4 illustrates the central ideas that we wish to analyze. Thus, in a flat manifold (Fig. 4a), two geodesics that are initially parallel, extended indefinitely, maintain their mutual distance constant; that is, the parallel postulate holds. In contrast, in a manifold with positive curvature two initially parallel geodesics converge (Fig. 4b), whereas in a manifold with negative curvature the geodesics diverge (Fig. 4c).\\

The above shows us a general result of great importance: a single geodesic does not allow us to determine the type of curvature of a manifold. For this purpose we need two geodesics; in particular, we need to determine whether the geodesics converge or diverge, that is, we must establish whether \textit{geodesic deviation} exists, which is a measure of the curvature of a manifold. The more pronounced the geodesic deviation, the greater the curvature. \\

\begin{figure}[h]
  \centering
    \includegraphics[width=0.25\textwidth]{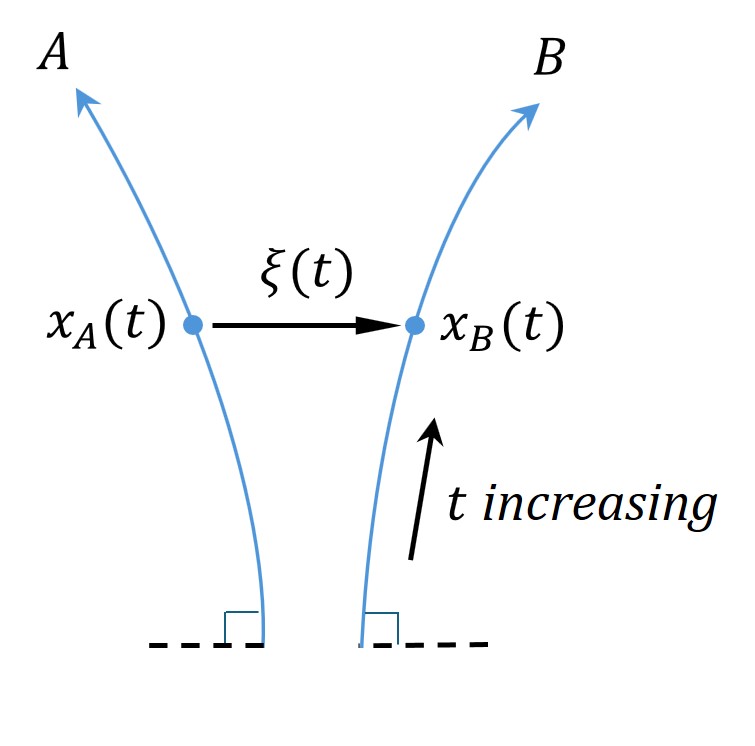}
  \caption{Geodesic deviation is measured from the separation $\xi (t)$ between two geodesics $x_{A}(t)$ and $x_{B}(t)$ whose mutual distance varies as a function of a parameter $t$.}
\end{figure} 

In Euclidean geometry, geodesic deviation is zero. In a non-Euclidean manifold the deviation is also zero locally, since, as we know, in a small neighborhood of a point, non-Euclidean manifolds agree with Euclidean geometry. Figure 5 illustrates these ideas, where two geodesics $A$ and $B$ are separated by a distance $\xi (t)$, with $t$ being a parameter that we again interpret as the time that two test particles take to move along their respective geodesics. If $x_{A}(t)$ and $x_{B}(t)$ are the positions of the particles in a given coordinate system, we can define intuitively the geodesic deviation as\footnote{It is important to keep in mind that this expression is only an intuitive representation in a chosen coordinate system, not the formal definition} [1,8]:

\begin{equation} %21
\xi(t) = x_{A}(t) - x_{B}(t).
\end{equation}

Thus, if $\xi(t)$ decreases over time, we are in the situation illustrated in Fig. 4(b), where the curvature is positive. If $\xi(t)$ increases, we are in the situation illustrated in Fig. 4(c), where the curvature is negative. In these two situations the manifold is non-Euclidean. But if $\xi(t)$ remains constant over time, we are in the situation illustrated in Fig. 4(a), where the curvature is zero and therefore the manifold is Euclidean.

\section{The Metric of a Non-Euclidean Manifold: A Simple Example}

The simplest example of non-Euclidean geometry is the surface of a sphere, whose curvature is positive and constant. To study this two-dimensional manifold, we must begin by finding the corresponding metric. First, it should be noted that a spherical surface of constant radius $r$ is described by two numbers, the polar angle $\theta$ and the equatorial angle $\varphi$, where $0 \leq \theta \leq \pi$ and $0 \leq \varphi \leq 2\pi$ (see Fig. 6). As illustrated, according to these conventions, the metric of the spherical surface can be written as

\begin{equation} %22
dl^{2}=r^{2}d\theta^{2} + r^{2} \sin^{2}\theta d\varphi^{2} \ (r \ constant).
\end{equation}

\begin{figure}[h]
  \centering
    \includegraphics[width=0.4\textwidth]{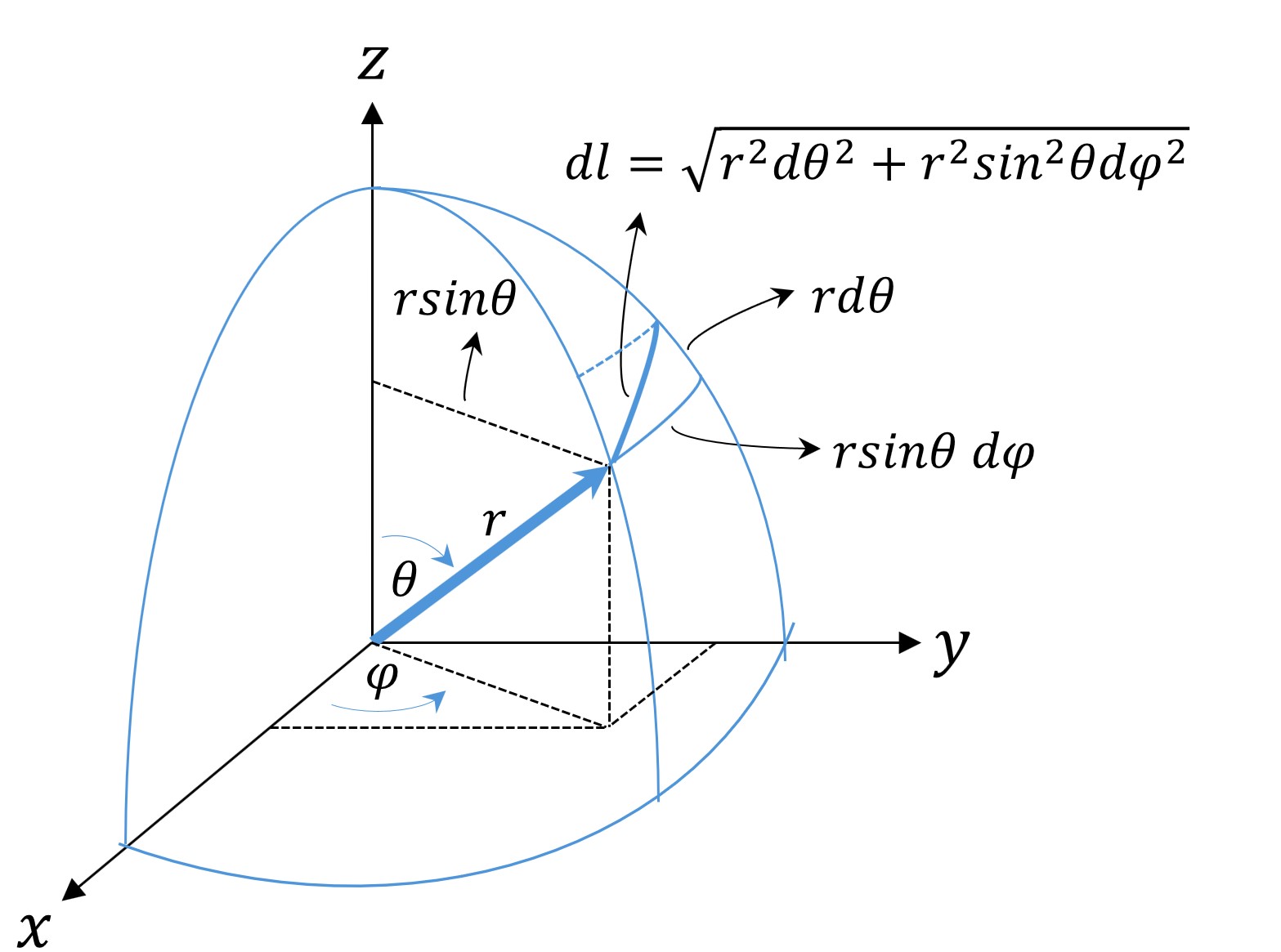}
  \caption{Polar angle $0 \leq \theta \leq \pi$, equatorial angle $0 \leq \varphi \leq 2\pi$, and the metric of the surface of a sphere.}
\end{figure} 

Therefore, the corresponding metric tensor is

\begin{equation} %23
\left[g_{ij} \right] = 
\begin{pmatrix}
r^{2} & 0\\
0 & r^{2} \sin^{2} \theta 
\end{pmatrix}.
\end{equation}

Although the spherical surface is curved, we have obtained this metric using the Pythagorean theorem of plane or Euclidean geometry, as required by the fundamental principle of Riemannian geometry.\\

Once the metric has been obtained, we must keep in mind that the basic criterion for determining whether a manifold is non-Euclidean is to find disagreements with some well-established result of Euclidean geometry, such as those illustrated in Fig. 7. To see how this criterion works and to practice the use of the metric, let us draw a circumference on the surface of a sphere, like the blue circumference in Fig. 8. We wish to calculate its perimeter $C$, for which we take $\theta=\text{constant}$ in Eq. (22), which implies that $d\theta=0$ and $\sin \theta=\text{constant}$. Then, integrating between $0$ and $2\pi$:

\begin{equation}%24
\int_{0}^{C} dl = r \sin\theta \int_{0}^{2\pi} d\varphi \rightarrow C=2\pi(r \sin\theta).
\end{equation}

Let us now determine the radius $\rho$ of the blue circumference, which we define as a geodesic segment between a point of the circumference and its center (the North Pole). In this case, $\varphi =\text{constant}$ and $d\varphi=0$ in Eq. (23). Integrating between $0$ and $\varphi$:

\begin{equation}%25
\int_{0}^{\rho} dl = r\int_{0}^{\theta} d\theta \rightarrow \rho =r\theta.
\end{equation}

Taking the ratio between the perimeter and the radius, given by Eqs. (24) and (25), we obtain

\begin{equation} %26
\frac{C}{\rho}=2\pi \left( \frac{\sin \theta}{\theta} \right) .
\end{equation}

We see that the recovery of flat Euclidean geometry is obtained locally by directly using the limit $\sin \theta/\theta \rightarrow 1$ as $\theta \rightarrow 0$, which satisfies the fundamental principle of Riemannian manifolds. It is easy to show (for example, graphically) that for $\theta>0$, $\sin \theta/ \theta <1$, so that

\begin{equation} %27
\frac{C}{\rho} < 2\pi.
\end{equation}

This result does not agree with what is established by Euclidean flat geometry, which tells us that the ratio between the perimeter and the radius of a circumference is equal to $2\pi$. We therefore conclude that the spherical surface is a non-Euclidean manifold. According to the concept of geodesic deviation introduced in the previous section, we can also arrive at this conclusion by noting that two initially parallel geodesics drawn on the spherical surface converge, as illustrated in Fig. 4(b), which reflects the fact that the curvature is positive.\\

\begin{figure}[h]
  \centering
    \includegraphics[width=0.6\textwidth]{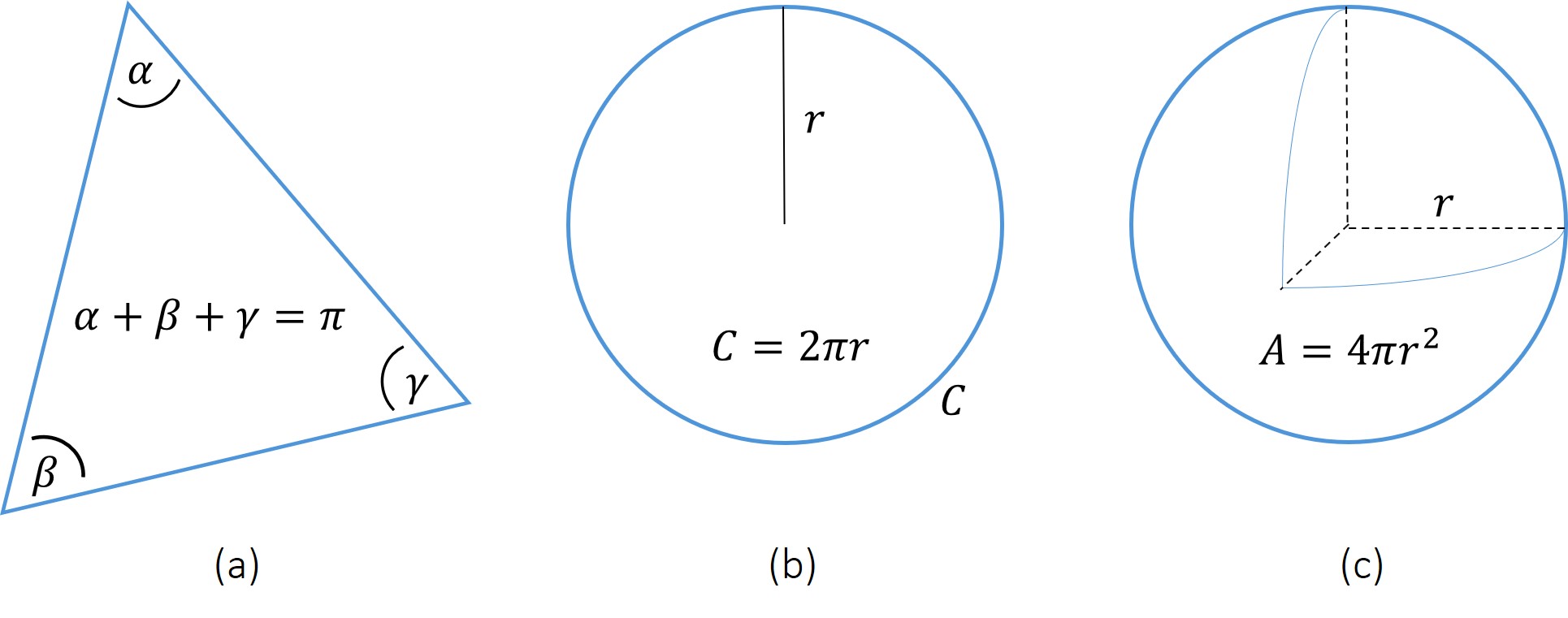}
  \caption{Some well-known results of Euclidean geometry.}
\end{figure} 

Continuing with the example of the spherical surface, we will conclude this section by analyzing a simple example of the fundamental principle of Riemannian geometry. To do this, let us return to Eq. (24), taking $\theta=\rho /r$:

\begin{equation}%28
C=2\pi r \sin\dfrac{\rho}{r}.
\end{equation}

We know that the Taylor series expansion of the function $\sin x$ is

\begin{equation}%29
\sin x = x-\dfrac{x^{3}}{3!} + O(x^{5}).
\end{equation}

Thus, Eq. (28) can be rewritten as

\begin{equation}%30
C=2\pi r\left( \frac{\rho}{r} - \dfrac{\rho^{3}}{3!r^{3}} + O(\rho^{5}) \right). 
\end{equation}

If we consider a very small region of the spherical surface, then $\rho\cong 0$ and therefore we can retain only the first term of the Taylor series, which gives the formula for the perimeter of a circumference in Euclidean geometry:

\begin{equation}%31
C = 2\pi r \left( \frac{\rho}{r} \right)=2\pi \rho.
\end{equation}

This result conﬁrms the validity of the fundamental principle of Riemannian geometry, since we see that locally, in a small region of the curved spherical surface, Euclidean flat geometry holds.

\begin{figure}[h]
  \centering
    \includegraphics[width=0.2\textwidth]{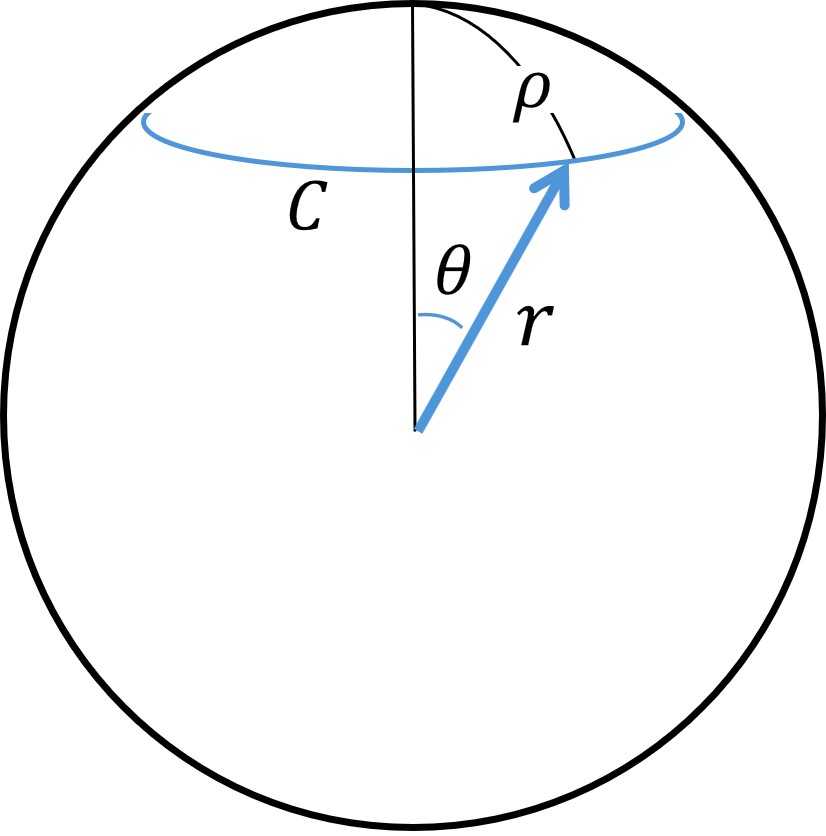}
  \caption{Circumference (blue) of perimeter $C$ and radius $\rho$ on the surface of a sphere of radius $r$.}
\end{figure}

\section{General Relativity, Roughly Speaking}

According to relativistic physics, time and the three spatial dimensions are intimately linked to form a four-dimensional continuum called \textit{space-time} [1]. As in Riemannian geometry, the fundamental object of this four-dimensional relativistic manifold is the metric, which defines the infinitesimal distance $ds$ between any two points in space-time.\\ 

Thus, Einstein took the concept of metric from Riemannian geometry\footnote{Whenever a squared inﬁnitesimal distance, that is an invariant homogeneous quadratic function of the coordinate diﬀerentials is deﬁned, we call the manifold a \textit{metric space} or a \textit{Riemannian space}.}, but introduced a subtle yet important modification by incorporating time as a new coordinate on equal footing with the spatial coordinates. The simplest relativistic metric, known as the \textit{Minkowski metric}, describes flat space-time in Cartesian coordinates [1]. If we denote by $c=3\times 10^{8} m\cdot s^{-1}$ the speed of light in vacuum, which is a fundamental constant of relativistic physics, the Minkowski metric can be written as [9]:

\begin{equation}%32
ds^{2}=c^{2}dt^{2}-dx^{2}-dy^{2}-dz^{2}=c^{2}dt^{2}-dl^{2},
\end{equation}

where $ds$ is known as the \textit{interval}, $cdt$ is the temporal separation between two infinitesimally close points in space-time, and $dx$, $dy$, $dz$ are the infinitesimal spatial separations, so that $dl^{2}= dx^{2}+dy^{2}+dz^{2}$ is the metric of a flat three-dimensional space in Cartesian coordinates, as defined in Riemannian geometry, Eq. (9). We see that, dimensionally, the temporal separation $cdt$ corresponds to a distance, so that Eq. (32) is dimensionally homogeneous. Just as $dl^{2}$ is an invariant in Riemannian geometry, $ds^{2}$ is a relativistic invariant, although its physical–geometrical interpretation is more subtle than in the Riemannian case. Nevertheless, it remains true that $ds^{2}$ is invariant and has the same value in all coordinate systems. \\ 

The part of the metric that contains the temporal coordinate $t$ is called the \textit{temporal part}, and the part containing the Cartesian coordinates $(x,y,z)$ is called the \textit{spatial part}. We can also write Eq. (32) in spherical coordinates (see Fig. 1), which will be useful later:

\begin{equation}%33
ds^{2} = c^{2}dt^{2} - dr^{2} - r^{2}d\theta^{2} - r^{2}\sin^{2}\theta d\varphi^{2} = c^{2}dt^{2}-dl^{2}.
\end{equation}

In this expression, the quantity

\begin{equation}%34
dl^{2} =  dr^{2} + r^{2}d\theta^{2} + r^{2}\sin^{2}\theta d\varphi^{2},
\end{equation}

is the infinitesimal distance between two points of flat three-dimensional space in spherical coordinates. Eqs. (32) and (33) confirm something we already know: the form taken by a metric depends on the coordinate system used. Using the concepts of Riemannian geometry, from Eq. (32) we see that the Minkowski metric tensor expressed in Cartesian coordinates can be written as

\begin{equation} %35
\left[ g_{\mu \nu} \right] = \begin{pmatrix} 1 & 0 & 0 & 0 \\ 0 & -1 & 0 & 0 \\0 & 0 & -1 & 0 \\ 0 & 0 & 0 & -1 \end{pmatrix}.
\end{equation}

To distinguish purely spatial metrics of Riemannian geometry from relativistic metrics, which are spacetime metrics, it is customary for the latter to use Greek subscripts such as $\mu$ and $\nu$ in Eq. (35). \\

Eqs. (32) and (33) show that $ds$ can be interpreted as a \textit{proper-time} interval, which we can intuitively define as the time measured by an observer using a wristwatch that they always carry with them. If we consider the coordinates of the observer at rest in their own frame of reference, where $dx=dy=dz=0$, we obtain

\begin{equation}%36
ds=cdt \equiv cd\tau,
\end{equation}

where $\tau$ is the conventional symbol used to denote proper time. This result has general validity, so that $ds=cd\tau$ always holds, both in special relativity and in GR. To distinguish the proper time $\tau$ from the time $t$ appearing on the right-hand side of the metric, the latter is called \textit{coordinate time}, and represents a time interval measured relative to a background coordinate system. Let us consider an object moving with instantaneous speed $v$; in a time interval $dt$ it travels the spatial distance $dl=vdt$, so that we can rewrite Eq. (32) in the form

\begin{equation}%37
ds^{2}=c^{2}dt^{2}-v^{2}dt^{2}=dt^{2}(c^{2}-v^{2}).
\end{equation}

Since relativistic physics requires that $v\leq c$, we see that two types of physically allowed intervals exist: the \textit{timelike interval} ($v<c$), which according to Eq. (37) satisfies the condition $ds>0$ and describes massive objects, and the \textit{lightlike interval} ($v =c$), which satisfies the condition

\begin{equation}%38
ds=0,
\end{equation}

and describes rays or particles of light whose rest mass is zero and whose speed is locally constant and equal to $c$ in all reference frames. For this reason, intervals described by photons or light rays are known as \textit{null geodesics}. This result also has general validity, so that $ds=0$ always holds, both in special relativity and in GR.\\ 

If we compare Eqs. (14) and (35), we see that there is great similarity between them, except for the negative sign of the coefficients in the spatial part. Because of this similarity, it is said that the Minkowski metric describes a \textit{pseudo-Euclidean} space-time. Geometrically, within relativistic physics this means that Eqs. (32) and (33) represent a flat four-dimensional space-time without curvature, just as Eq. (14) describes a flat four-dimensional space. This flat space-time is the stage on which the phenomena of the theory of special relativity occur, which does not incorporate gravity. If we consider two test particles describing initially parallel geodesics in this scenario, they will maintain their mutual distance constant through time, and therefore their trajectories in space-time will be analogous to those illustrated in Fig. 4a. \\

How is a geodesic defined in space-time? If a test particle moves freely under the action of gravity, we say that it describes a \textit{spacetime geodesic}. When we speak of free motion we mean that no force acts on the particle, including gravity, as noted in the introduction, within the framework of GR gravity is not a force but a manifestation of space-time curvature. \\

Thus, by definition, the trajectory of a free particle is a spacetime geodesic and, as in Riemannian geometry, it can be defined as the locally straightest path between two infinitesimally close points. This definition is supported by the \textit{weak equivalence principle} [2], a result confirmed by numerous experiments, which states that all objects in free fall have the same acceleration regardless of their composition or structure. This reveals that the effects of gravity depend on a property residing in gravity itself and not in the objects affected by it. This property is the curvature of space-time.\\ 

To explore the concept of spacetime curvature more deeply, imagine two test particles moving freely through a region without gravity, following initially parallel trajectories. By definition, the trajectories are spacetime geodesics. Figure 9a shows the particles moving through space. Figure 9b shows the motion in a spacetime diagram, where it can be seen that the mutual distance $\xi(t)$ remains constant, that is, there is no geodesic deviation, which indicates that the curvature of space-time is zero, just as occurs on a flat surface illustrated in Fig. 9c, which represents flat Minkowski spacetime. \\

\begin{figure}[h]
  \centering
    \includegraphics[width=0.7\textwidth]{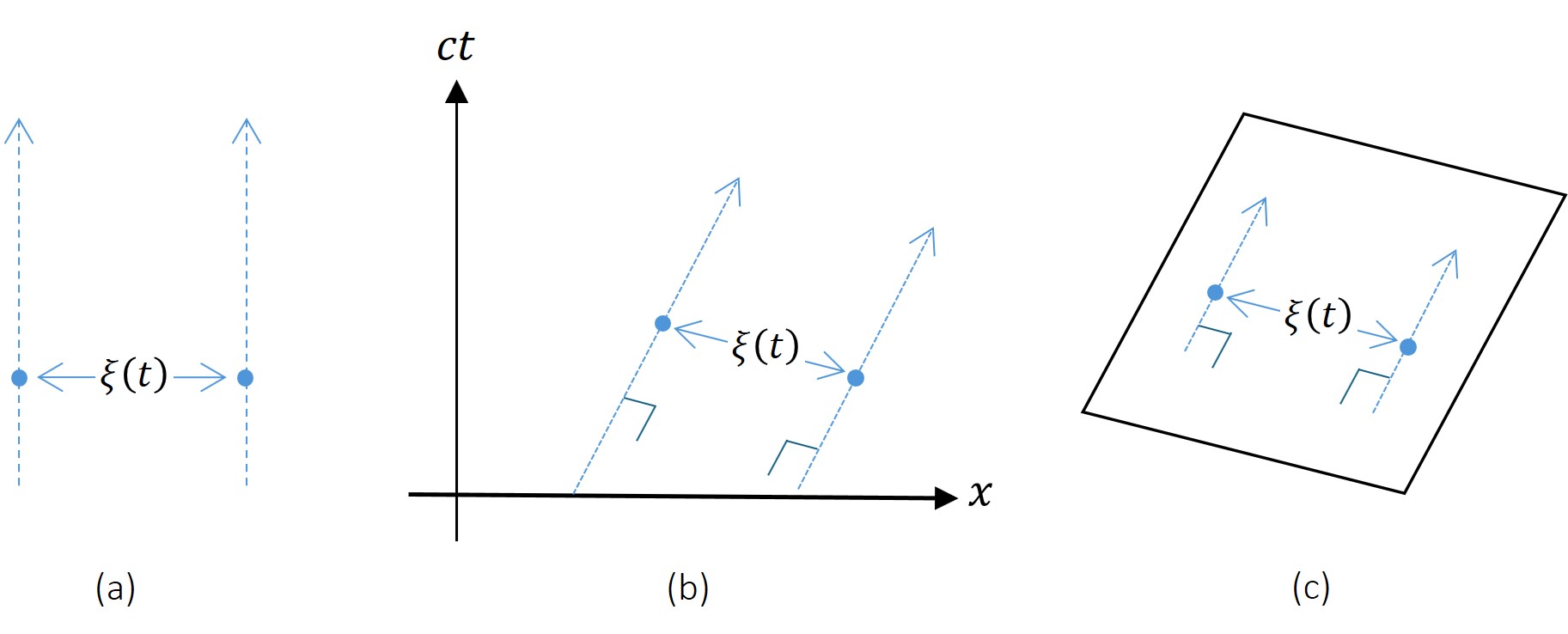}
  \caption{Illustration of zero spacetime curvature in an arbitrary $x$ direction.}
\end{figure} 

In perfect analogy with the ideas discussed in previous sections, it can be shown that when curvature is zero at every point of space-time, it is always possible to find a coordinate system where the metric tensor takes the simple diagonal form given by Eq. (35).\\ 

If we now place the two test particles in a region where there is a gravitating body such as the Earth, it is clear that $\xi(t)$ cannot remain constant, since the particles will be deflected by gravity, possibly approaching each other in some regions and moving apart in others. In other words, gravity produces geodesic deviation, so that space-time curves in the presence of a distribution of mass or its equivalent in energy. Figure 10 illustrates a situation where the particles approach one another. If the particles are released from a certain height above the Earth's surface (Fig. 10a), they will follow spacetime geodesics. As a result, $\xi(t)$ will decrease, and if we plot the trajectories of the particles in a spacetime diagram we will see that they are initially parallel and then curve toward one another (Fig. 10b), in a manner analogous to what occurs with geodesics on the surface of a sphere (Fig. 10c). The fact that the geodesics are initially parallel is due to the fact that the initial speed of the particles is zero, so the geodesics are instantaneously parallel to the $ct$ axis. Thus we may say that spacetime curvature is positive in the transverse $\theta$ direction relative to the radius.\\ 

\begin{figure}[h]
  \centering
    \includegraphics[width=0.7\textwidth]{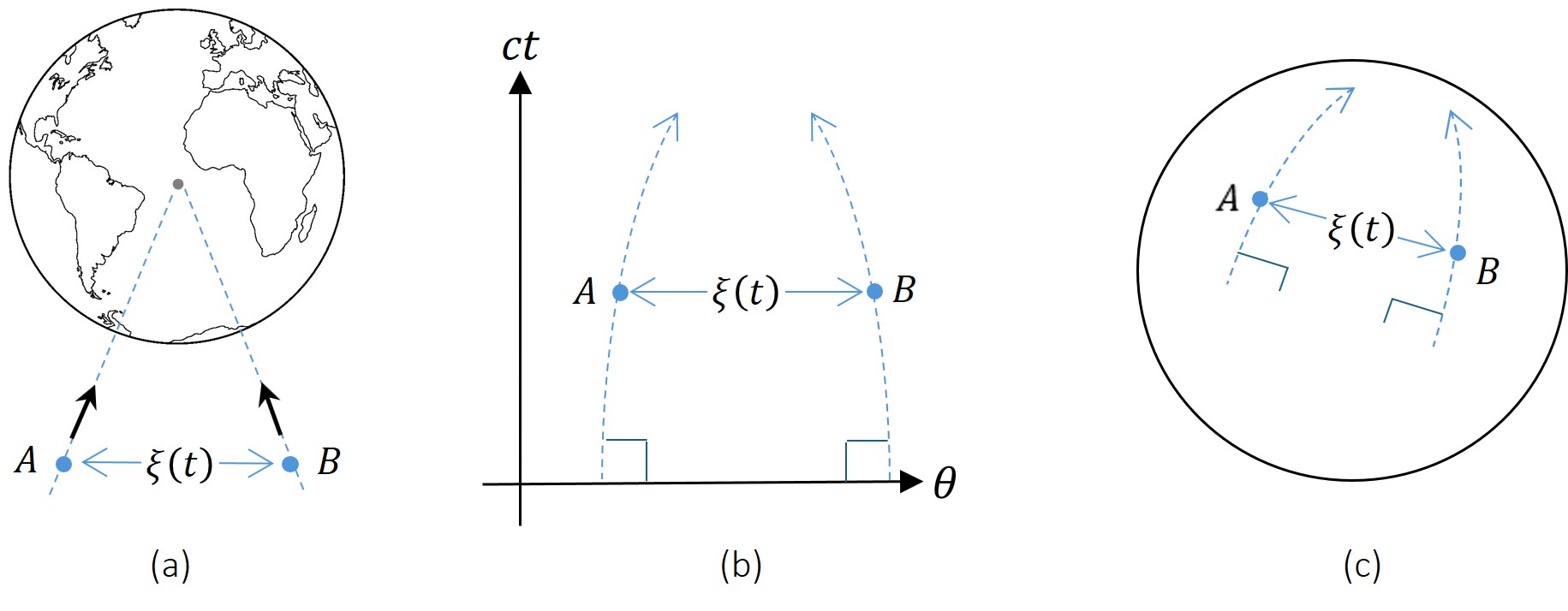}
  \caption{Illustration of positive spacetime curvature in the transverse $\theta$ direction.}
\end{figure} 

Formally, spacetime curvature is manifested by the presence of metric components that depend on the distribution of mass. This means that special relativity is a particular case of GR for situations where the mass-energy distribution is zero or so small that it can be neglected. Consequently, the fundamental principle of Riemannian geometry is also valid here, which states that locally, in a curved manifold, Euclidean geometry holds. We may therefore say that locally, in a small region of curved space-time, special relativity holds, since locally space-time is flat to first approximation. \\ 

However, this last statement presents a delicate physical problem, since special relativity is strictly valid only in inertial reference frames where gravity is absent, and a particle falling freely toward a celestial body experiences gravity. Einstein overcame this difficulty by noting that an observer in free fall feels weightless, as if floating in a region of space where no gravitating masses exist. The great physicist embodied this idea in the equivalence principle: \textit{A freely falling reference frame in a gravitational field is locally equivalent to an inertial frame (without gravity)} [10]. Thus, the equivalence principle guarantees that in a small neighborhood of a point in curved space-time, special relativity agrees with GR. This means that a test particle moving freely through space-time locally does not experience gravity and follows a locally straight geodesic. By integrating these straight geodesics we obtain the curved spacetime geodesics. This important result is responsible for the fact that Eqs. (36) and (38) are valid both in special and general relativity.\\ 

GR is expressed through a set of 16 coupled partial differential equations that relate spacetime curvature to the distribution of mass-energy. These equations can be written compactly through a single tensorial expression known as the \textit{Einstein equation}\footnote{This is Einstein’s equation without the cosmological constant, a quantity added by the physicist in 1917 to produce a model of a static universe consistent with his worldview. Einstein’s cosmological constant generates a repulsive effect that counteracts gravity, producing a universe that maintains constant size, so that it neither expands nor contracts. When Hubble discovered the expansion of the universe in 1930, Einstein referred to the cosmological constant as the greatest “blunder” of his scientific career.} [6,11,12]: 

\begin{equation} %39
R_{\mu \nu} - \frac{1}{2}Rg_{\mu \nu} = \frac{8\pi G}{c^{4}} T_{\mu \nu},
\end{equation}

where $G=6,67 \times 10^{-11} N\cdot m^{2} \cdot kg^{-2}$ is the gravitational constant, $c$ is the speed of light in vacuum, $g_{\mu \nu}$ is the metric tensor discussed in the previous section, $T_{\mu \nu}$ is the \textit{energy–momentum tensor}, $R$ is the \textit{Ricci scalar}, and $R_{\mu \nu}$ is the \textit{Ricci tensor}, which involves first and second-order derivatives of the metric tensor, making Einstein’s equations a nonlinear system of coupled partial differential equations. Since this equation establishes a relation between tensors, it is covariant, a property that Einstein considered fundamental for formulating a physical law. Although a detailed analysis of Einstein’s equation goes beyond the scope of this work, we can extract some of its physical meaning by noting that the left-hand side ($R_{\mu \nu}$, $R$, and $g_{\mu \nu}$) describes the curvature of space-time, whereas the right-hand side ($T_{\mu \nu}$) describes the mass–energy density responsible for that curvature. Symbolically we may write Eq. (39) as

\begin{equation} %40
\left(  \begin{array}{c} space\text{-}time \\ curvature \end{array} \right) = \frac{8\pi G}{c^{4}}\left( \begin{array}{c} mass\text{-}energy \\ distribution \end{array} \right).
\end{equation}

The central idea underlying Einstein’s equation can be summarized by recalling a famous phrase by the physicist John A. Wheeler: \textit{spacetime tells matter how to move, and matter tells spacetime how to curve} [7]. In other words, within the framework of GR gravity is pure geometry, and if we know the distribution of mass-energy in a region of space we know its spacetime geometry. A simple particular case occurs when the mass-energy distribution is zero and the boundary conditions select flat spacetime. In that case, Einstein’s equations admit the Minkowski solution, for which spacetime curvature vanishes:

\begin{equation} %41
\left(  \begin{array}{c} space\text{-}time \\ curvature \end{array} \right) = 0.
\end{equation}

More generally, however, vacuum solutions ($T_{\mu \nu}=0$) do not necessarily imply vanishing curvature. For example, the exterior Schwarzschild solution satisfies $R_{\mu \nu}=0$ while still possessing nonzero spacetime curvature associated with tidal gravitational effects. In this case there is a trivial solution to Einstein’s equation which is given by the metric components $g_{00}=1$, $g_{11}=g_{22}=g_{33}=-1$, and $g_{\mu \nu}=0$ if $\mu \neq \nu$, so that the metric tensor takes the Minkowskian form given by Eq. (35) (see Appendix B).\\ 

The subscripts $\mu$ and $\nu$ in Eq. (39) take the values $0,1,2,3$, where by convention zero denotes the temporal coordinate and the other numbers denote the three spatial coordinates. Since we can form 16 pairs of numbers with $0,1,2,3$, Einstein’s equation contains 16 equations. To appreciate the extraordinary power of tensor notation, which allowed Einstein to package 16 equations into a single mathematical formula, it is useful to write this equation in matrix form:

\begin{equation} %42
\begin{pmatrix} R_{00} & R_{01} & R_{02} & R_{03} \\ R_{10} & R_{11} & R_{12} & R_{13} \\ R_{20} & R_{21} & R_{22} & R_{23} \\ R_{30} & R_{31} & R_{32} & R_{33} \end{pmatrix} - \dfrac{1}{2}R \begin{pmatrix} g_{00} & g_{01} & g_{02} & g_{03} \\ g_{10} & g_{11} & g_{12} & g_{13} \\ g_{20} & g_{21} & g_{22} & g_{23} \\ g_{30} & g_{31} & g_{32} & g_{33} \end{pmatrix} = \dfrac{8\pi G}{c^{4}} \begin{pmatrix} T_{00} & T_{01} & T_{02} & T_{03} \\ T_{10} & T_{11} & T_{12} & T_{13} \\ T_{20} & T_{21} & T_{22} & T_{23} \\ T_{30} & T_{31} & T_{32} & T_{33} \end{pmatrix}.
\end{equation}

Each of these 16 equations is obtained by equating matrix elements with identical subscripts, for example:

\begin{equation} %43
R_{01} - \frac{1}{2}R g_{01} = \dfrac{8\pi G}{c^{4}} T_{01},\qquad
R_{23} - \frac{1}{2}R g_{23} = \dfrac{8\pi G}{c^{4}} T_{23}.
\end{equation}

The “unknown” in Eq. (39) is the metric tensor $g_{\mu \nu}$. Thus, the central objective when solving Einstein’s equation is to find the components $g_{\mu \nu}$ of the corresponding matrix. As in Riemannian geometry, the importance of these components lies in the fact that they allow us to construct the metric, which provides the infinitesimal distance between any two points of space-time. For a given mass–energy distribution, the metric contains all the physical and geometrical information about the structure of space-time [4].\\ 

In practice, Einstein’s equation contains 10 independent equations rather than 16, since not only the metric tensor but also the Ricci tensor and the energy–momentum tensor are symmetric, meaning that $g_{\mu \nu}= g_{\nu \mu}$, $R_{\mu \nu}= R_{\nu \mu}$, and $T_{\mu \nu}= T_{\nu \mu}$. \\ 

Most metrics of interest in GR can be written in diagonal form, meaning that they contain nonzero components $g_{\mu \nu}$ only when $\mu=\nu$. The Schwarzschild metric is also of this type, so that for our purposes the metric of a four-dimensional spacetime can be written generically as

\begin{equation} %44
ds^{2}= g_{\mu \nu}dx_{\mu}dx_{\nu} = g_{00}dx_{0}^{2}+g_{11}dx_{1}^{2}+g_{22}dx_{2}^{2}+g_{33}dx_{3}^{2},
\end{equation}

where $g_{00}$ is the component of the metric tensor corresponding to the temporal part, and $g_{11}$, $g_{22}$, $g_{33}$ are the components of the spatial part. An important feature of this metric —and in general of any well-defined spacetime metric— is that the temporal part always has a sign opposite to that of the spatial part. It does not matter whether the minus sign is placed in front of the temporal or spatial part, since this is merely a convention that does not alter the properties of the spacetime geometry considered. This convention holds both in special relativity and in GR. To account for this convention the concept of \textit{signature} is used [11], and we say, for example, that the signature of Eqs. (32) and (33) is

\begin{equation} %45
sig = \lbrace + - - - \rbrace.
\end{equation}

However, the signature of Eqs. (32) and (33) can also be written equivalently as

\begin{equation} %46
sig = \lbrace - + + + \rbrace,
\end{equation}

which allows us, for example, to rewrite Eq. (32) as

\begin{equation}%47
ds^{2}=-c^{2}dt^{2} +dx^{2}+dy^{2}+dz^{2}=-c^{2}dt^{2}+dl^{2}.
\end{equation}

The concept of signature has no meaning in Riemannian geometry, since Riemannian metrics are always positive definite; that is, all the metric components are greater than or equal to zero. However, this is not the case for the metrics studied in GR, where both negative and positive coefficients appear. For this reason, spacetime geometries are called \textit{pseudo-Riemannian}, and in particular, as mentioned earlier, the Minkowski metric is said to be pseudo-Euclidean. From now on we will preferentially, though not exclusively, use the signature defined in Eq. (45).

\section{The Exterior Schwarzschild Solution}

Solving Einstein’s equation means finding the metric for a given distribution of mass–energy in space-time. However, Einstein’s equation is too complex to solve in full generality. In many situations it is sufficient to find approximate solutions, which is what Einstein himself did when he formulated his equation.\\ 

\begin{figure}[h]
  \centering
    \includegraphics[width=0.3\textwidth]{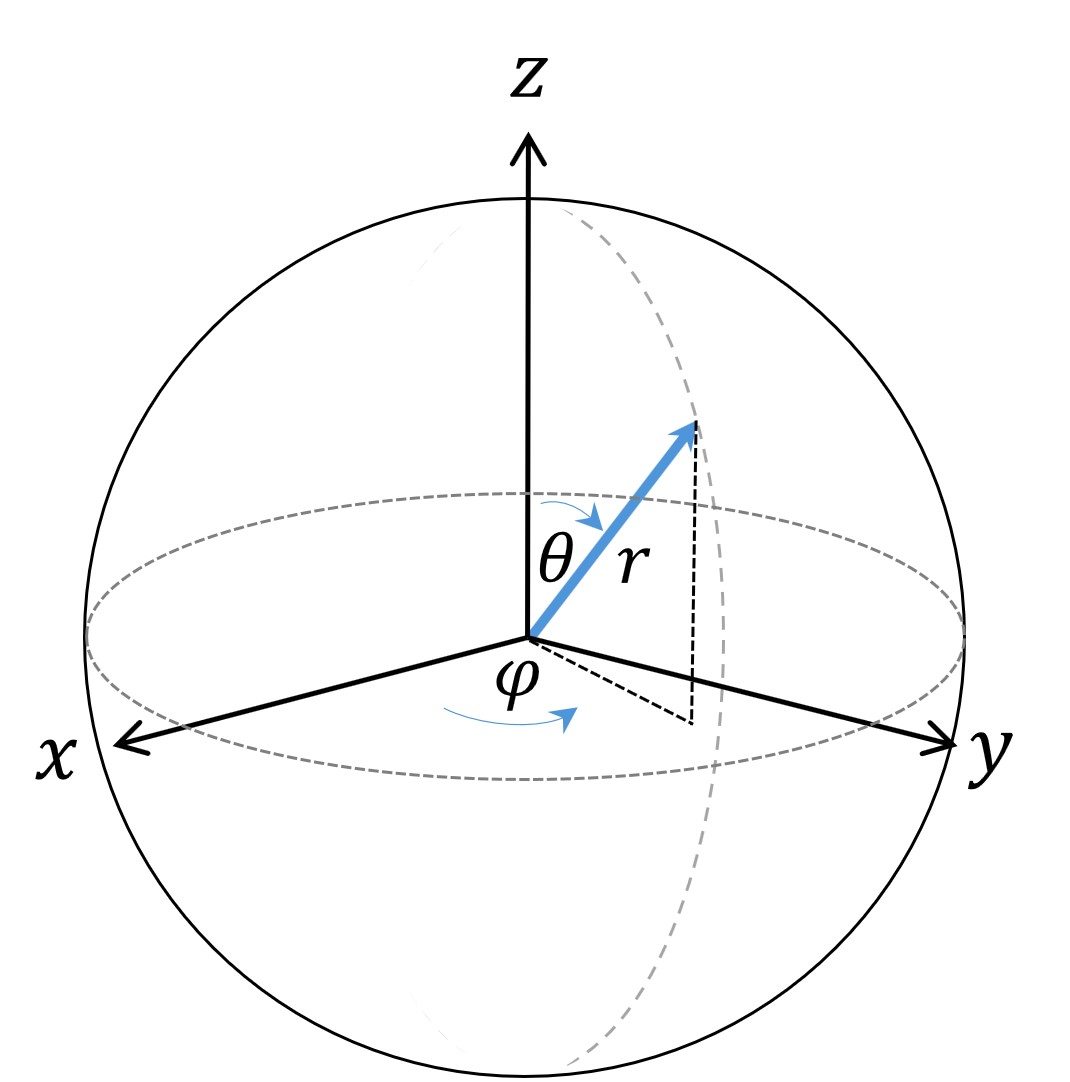}
  \caption{Spherical coordinates, $r>0$, $0 \leq \theta \leq \pi$, and $0 \leq \varphi \leq 2\pi$.}
\end{figure} 

Several exact solutions are known that describe space-times with symmetry conditions sufficiently restrictive for Eq. (39) to be simplified and solved without approximations [13]. However, only a few are of interest for physics and astronomy. One of them is the Schwarzschild solution, or \textit{Schwarzschild metric} [14-16]. This solution describes a static and stationary spacetime in the exterior region of a non-rotating, isolated spherically symmetric body of mass $M$ and radius $R$, such as a star or a planet. This implies calculating the components $g_{00}, g_{11}, g_{22}, g_{33}$ for $r>R$ in the vacuum space-time\footnote{Schwarzschild also found an exact solution to Einstein’s equation for the space-time inside a massive body with spherical symmetry.}. That is, for $r>R$ no matter exists, which allows us to set $T_{\mu\nu}=0$ in Eq. (39). Consequently, the Schwarzschild metric is obtained by solving the reduced Einstein equation

\begin{equation}%48
R_{\mu\nu} -\frac{1}{2}Rg_{\mu\nu}=0.
\end{equation}

Using the so-called \textit{tensor calculus} it can be shown that, under the conditions imposed by Schwarzschild, Eq. (48) reduces to $R_{\mu\nu}=0$ (Appendix C). However, the vacuum equations alone do not uniquely determine the spacetime geometry. Additional physical boundary conditions are required in order to specify the solution. In the Schwarzschild case, the parameter $M$ is fixed by demanding that the metric reproduce Newtonian gravitation in the weak-field limit far from the source. Furthermore, because of the symmetry of the problem it is possible to use spherical coordinates (Fig. 11), which implies that 

\begin{equation}%49
x_{0}=ct, x_{1}=r,\ x_{2}=\theta,\ x_{3}=\varphi.
\end{equation}

These are the so-called \textit{Schwarzschild coordinates}. Expressed in these coordinates, the general form of the Schwarzschild metric is

\begin{equation} %50
ds^{2}= g_{00}c^{2}dt^{2}+g_{11}dr^{2}+g_{22}d\theta^{2}+g_{33}d\varphi^{2}.
\end{equation}

After considerable mathematical work it is found that, under the conditions imposed above, the Schwarzschild metric tensor can be written as

\begin{equation} %51
\left[ g_{\mu \nu} \right]  = \begin{pmatrix} 1-\frac{2GM}{c^{2}r} & 0 & 0 & 0 \\ 0 & -\left( 1-\frac{2GM}{c^{2}r} \right)^{-1}  & 0 & 0 \\ 0 & 0 & -r^{2} & 0 \\ 0 & 0 & 0 & -r^{2} \sin^{2}\theta \end{pmatrix}.
\end{equation} 

where

\begin{equation}%52
g_{00} =1-\frac{2GM}{c^{2}r},\quad 
g_{11}=-\left( 1-\frac{2GM}{c^{2}r} \right)^{-1},\quad 
g_{22}=-r^{2},\quad 
g_{33}=-r^{2} \sin^{2}\theta,\quad 
g_{\mu\nu}=0 \ \text{if}\ \mu \neq\nu.
\end{equation}

Therefore, the Schwarzschild solution takes the form [12,13]

\begin{equation} %53
ds^{2} = c^{2}\left( 1 - \frac{2GM}{c^{2}r} \right)dt^{2} - \left( 1 - \frac{2GM}{c^{2}r} \right)^{-1}dr^{2} - r^{2}d\theta^{2} - r^{2}\sin^{2}\theta d\varphi^{2}.  
\end{equation}

The presence of coefficients that depend on $M$ reveals that this is the metric of a space-time with gravity and therefore curved. This equation is also written as

\begin{equation} %54
ds^{2} = c^{2}\left( 1 - \frac{R_{S}}{r} \right)dt^{2} - \left( 1 - \frac{R_{S}}{r} \right)^{-1}dr^{2} - r^{2}d\theta^{2} - r^{2}\sin^{2}\theta d\varphi^{2},  
\end{equation}

where the quantity

\begin{equation}%55
R_{S}= \frac{2GM}{c^{2}}
\end{equation}

is called the \textit{Schwarzschild radius}, and provides a measure of the size of a \textit{Schwarzschild black hole}, that is, a non-rotating and uncharged black hole described by the Schwarzschild solution [14]. This, however, was understood decades after Schwarzschild found his solution. The black hole described by the Schwarzschild metric is the simplest one, since its mathematical description depends only on $M$ [15]. The Schwarzschild radius defines a spherical surface called the \textit{event horizon}, or simply the \textit{horizon}, which has no material existence but can be imagined as a one-way membrane that allows the flow of matter and energy only toward the interior. That is, the horizon determines the point of no return, from which no form of matter or energy can escape the gravity of the black hole [4]. \\

The metric contains all the information about Schwarzschild space-time [6]. As we will see in Section 9 through four simple examples, this information is extracted by mathematically manipulating the Schwarzschild metric. This is analogous, though more complex, to what we did in Section 4 by manipulating the metric of the surface of a sphere.\\

If we assume that all the mass described by the metric is concentrated at $r=0$, then Eq. (53) represents the metric of a Schwarzschild black hole. As the reader may have noticed, at $r=0$ the component $g_{00}$ of the metric becomes infinite, which indicates that this is a \textit{singular point}. It is also said that a \textit{singularity} exists at $r=0$. In this specific case it is a \textit{physical singularity}, meaning that it cannot be removed by any mathematical procedure [14]. The Schwarzschild metric also has a singularity at $r=R_{S}$, where the component $g_{11}$ becomes infinite. It can be shown, however, that in this case it is a \textit{mathematical singularity}, meaning that it can be removed by a simple coordinate transformation [14] \footnote{As follows from this analysis, the presence of a physical singularity is defined based on invariant criteria that do not exist in the case of a mathematical singularity. Therefore, the singularity at $r=0$ cannot be diagnosed simply because the component $g_{00}$ becomes infinite, since additional criteria are required to characterize singular behavior, whose detailed analysis lies beyond the scope of this work.} . This point becomes clearer when the metric is expressed in Kruskal–Szekeres coordinates. \\

From the metric tensor in general, and from the Schwarzschild metric in particular, we can calculate geodesics in GR by a procedure analogous to that used in Riemannian geometry. Recall that in curved space-time gravity is not a force but a manifestation of the curvature of space-time. Therefore, a particle moving solely under the action of gravity moves freely, and its motion is determined only by the curved spacetime geometry. For example, a particle released from a certain height above the surface of the Earth follows a geodesic while falling. The same is true of the trajectory of a projectile in the absence of non-gravitational forces, whose parabolic motion is a geodesic. Similarly, a planet following an elliptical orbit also traces a spacetime geodesic as it revolves around its star.\\ 

These ideas allow us to generalize Newton’s first law of motion by stating that a particle on which no force acts will move in a straight line if space-time is flat, and in general will move along curved spacetime geodesics in regions where gravitating bodies exist. In Section 9 we will analyze very simple examples of geodesic calculations in the cases of a material particle and of a photon.

\section{The Newtonian Limit of the Schwarzschild Metric}

When gravity is weak and the curvature of space-time is small, it can be shown that Einstein’s equation reduces to the \textit{Poisson equation} [12]:

\begin{equation}%56
\nabla^{2}\varphi = 4\pi G\rho,
\end{equation}

which is the most general form that Newton’s law of universal gravitation can take, where $\varphi$ is the gravitational potential, $\rho$ is the mass density, $\nabla$ is the \textit{nabla operator}, and $\nabla^{2}$ is the square of this operator, known as the \textit{Laplacian}\footnote{Nabla is a linear differential operator that acts on vector fields defined on a differentiable manifold. In Cartesian coordinates, the nabla operator is defined as $\nabla \equiv \hat x \frac{\partial}{\partial x}+ \hat y \frac{\partial}{\partial y} + \hat z \frac{\partial}{\partial z}$. Therefore, the Laplacian operator, which is the square of this operator, is defined as $\nabla^{2} \equiv \frac{\partial^{2}}{\partial x^{2}}+ \frac{\partial^{2}}{\partial y^{2}} + \frac{\partial^{2}}{\partial z^{2}}$.}. In analogy with Eq. (39), the left-hand side of the Poisson equation represents gravity (curvature), and the right-hand side represents the mass (or energy) distribution. Recall that according to the law of universal gravitation, the gravitational field generated by a uniform spherical body of mass $M$ at a distance $r$ from its center is written as

\begin{equation}%57
\vec{g}(r) = - \frac{GM}{r^{2}}\hat r = -\nabla\varphi (r), 
\end{equation}

where $\hat r$ is a unit vector pointing in the direction of $r$, and $\varphi (r)$ is the corresponding gravitational potential. There are several mathematical routes to show that in the limit of weak gravitational fields the Schwarzschild metric reduces to Newtonian gravitation. Here we will use a simple result demonstrated by Rindler in Chapter 9 of [17], which states that the temporal component $g_{00}$ of the Schwarzschild metric can be written in an alternative form as

\begin{equation}%58
g_{00}= e^{2\varphi(r)/c^{2}}.
\end{equation}

On the other hand, it can be shown that when gravity is weak and the curvature of space-time is small, the dominant metric coefficient in Eq. (53) is the temporal coefficient $g_{00}$, and the other three coefficients are needed only to describe extreme phenomena such as, for example, black holes. Comparing Eqs. (53) and (58) we obtain

\begin{equation}%59
e^{2\varphi(r)/c^{2}}= 1 - \frac{2GM}{c^{2} r}.
\end{equation}

Therefore,

\begin{equation}%60
\varphi(r)=\frac{c^{2}}{2}\ln\left( 1 - \frac{2GM}{c^{2} r} \right).
\end{equation}

This expression can be expanded in a Taylor series:

\begin{equation}%61
\varphi(r)= -\frac{GM}{r} + \frac{1}{c^{2}}\frac{(GM)^{2}}{r^{2}} - O\left( \frac{1}{r^{3}} \right). 
\end{equation}

For large distances (but not infinite) from the central mass $M$, the term $2GM/c^{2} r$ in Eq. (60) is very small. Consequently, we can keep only the first term of the expansion, obtaining the Newtonian gravitational potential:

\begin{equation}%62
\varphi(r)\cong -\frac{GM}{r}. 
\end{equation}

As we know, this potential produces Newton’s law of universal gravitation, since the potential is related to the gravitational field $\vec{g}(r)$ through Eq. (57). \\

In summary, Newtonian gravitation describes a space-time so slightly curved that it is almost flat. This explains, for example, why in the solar system —where the Sun’s gravity is comparatively weak— Newton’s law of gravitation is an excellent approximation, making it possible to perform very precise calculations that describe everything from planetary orbits to the trajectories of spacecraft.

\section{Some Properties of the Schwarzschild Metric}

Apart from its spherical symmetry, there are three other important properties of Eqs. (53) and (54), two of which were briefly mentioned earlier.\\

The first property is that the metric is \textit{stationary}, which means that the structure of Schwarzschild spacetime is independent of time and looks exactly the same at any instant [4]. In other words, the metric is invariant under a translation in time. Indeed, by examining matrix (51) we see that the variable $t$ does not appear in any of the components. Formally, we can prove that the metric is stationary by performing the transformation $t \rightarrow t' = t + t_{0}$, where $t_{0}$ is a constant. This transformation corresponds to a translation in time. Taking differentials on both sides of the equality eliminates the constant, which implies that $dt = dt'$.\\

The second property is that the metric is \textit{static}, which means that the structure of Schwarzschild spacetime does not distinguish between past and future, so that the metric is invariant under the transformation $t \rightarrow -t$ [4]. Thus, by taking differentials and squaring we obtain $(-dt)^{2} = dt^{2}$. \\

An important result related to the previous two properties is the \textit{Birkhoff--Jebsen theorem}, which states that every spherically symmetric vacuum solution of Einstein’s equations is necessarily stationary and described by the Schwarzschild metric [4]. In this sense, the theorem may be regarded as the relativistic analogue of Gauss’s theorem in Newtonian gravitation, since it implies that the exterior gravitational field depends only on the total mass of the source and not on its internal radial dynamics\footnote{In Newtonian gravitation, Gauss’s theorem states that the gravitational flux through any closed surface depends only on the total mass enclosed by that surface. In mathematical form, $\oint \vec{g}\cdot d\vec{A} = -4\pi GM$, where $\vec{g}$ is the gravitational field, $d\vec{A}$ is an infinitesimal outward-oriented area element, $G$ is the gravitational constant, and $M$ is the enclosed mass. For a spherically symmetric mass distribution, this theorem implies that the exterior gravitational field is identical to that produced by a point mass located at the center.} . In physical terms, this means that the exterior gravitational field of a spherically symmetric body remains Schwarzschild even if the body undergoes radial oscillations, provided the spherical symmetry is preserved. \\

The third property is that the metric is \textit{asymptotically flat}, which means that spacetime far from a gravitating body is nearly flat and, as we already know, Newton’s law of gravitation is a good approximation [18]. Indeed, by taking $r \rightarrow \infty$ in Eq. (53) we recover Eq. (33), which is the Minkowski solution in spherical coordinates for a flat and empty spacetime, where the curvature is zero.\\

This allows us to identify $t$ and $r$ as the temporal and radial coordinates for a very distant observer. That is, only at a large distance from $r = R_{S}$ can an observer attribute to the variables $t$ and $r$ the intuitive meaning they have in our everyday experience [11]. However, near the Schwarzschild radius, $t$ and $r$ are merely coordinates that describe spacetime and do not have a direct physical meaning. When we introduce the embedding diagrams in Section 10 we will analyze this idea intuitively for the case of the radial coordinate $r$.\\

Finally, we see that for $R_{S}/r \ll 1$ ($r \gg R_{S}$), the Schwarzschild metric also reduces to the metric of a flat spacetime, which is given by Eq. (33), and therefore the coordinates $t$ and $r$ have an intuitive physical meaning. This is the typical situation for most celestial bodies, whose radii $R$ are very large compared with their corresponding Schwarzschild radii [4]. As the reader can easily verify, for the Sun and the planets of the solar system it holds that $R_{S}/r \ll 1$, which implies that spacetime beyond the surface of these celestial bodies is almost flat. \\

For this reason, in most situations of astronomical interest, the predictions of GR are practically indistinguishable from those of Newton’s law of gravity. As an example, the radius of the Earth is $r_{\oplus} \sim 10^{6} m$, and its Schwarzschild radius is $R_{S} \sim 10^{-2} m$, so that $R_{S}/r_{\oplus} \sim 10^{-8} \ll 1$. The situation is different for the so-called \textit{compact objects}, such as neutron stars and black holes, for which $R_{S}/r \approx 1$, and therefore GR is required to describe them in detail.

\section{Four Simple Applications of the Schwarzschild Metric}

The applications of the Schwarzschild metric are broad and varied, and analyzing them in detail goes beyond the purposes of this work. Some of these applications require the use of very sophisticated mathematical tools and are therefore not suitable for an introductory article. Below we analyze four applications that, while satisfying the requirement of simplicity, are sufficiently interesting and varied for the reader to gain an idea of the power and usefulness of the Schwarzschild metric as a tool for describing gravitational and astronomical phenomena.

\subsection{Gravitational Time Dilation}

Let us consider a point in space located at a constant distance from the center of a spherical gravitating body of uniform mass $M$ (for example, a star). This is equivalent to saying that the point is at rest with respect to the Schwarzschild coordinates. Under these conditions we have $dr = d\theta = d\varphi = 0$, and the Schwarzschild metric reduces to

\begin{equation}%63
ds= c\left( 1- \frac{R_{S}}{r} \right)^{1/2} dt.
\end{equation}

Introducing Eq. (36) and integrating we obtain

\begin{equation}%64
\Delta \tau= \left( 1- \frac{R_{S}}{r} \right)^{1/2}\Delta t,
\end{equation}

where $\Delta \tau$ is the proper time measured by a clock located at a distance $r$ from the gravitating body, and $\Delta t$ is the coordinate time measured by a clock at infinity. We observe that $\Delta \tau < \Delta t$, that is, the clock measuring the proper time runs slower than the distant clock. This effect is called \textit{gravitational time dilation}, and it can also be formulated in terms of two clocks located at a finite distance from the spherical gravitating body. \\

To do this, let us consider two clocks, one of which is at a distance $r_{1}$ from the gravitating body and measures a proper time $\Delta \tau_{1}$, and the other is located at a distance $r_{2} < r_{1}$ from the body and measures a proper time $\Delta \tau_{2}$. Then, from Eq. (64) we have

\begin{equation}%65
\Delta \tau_{1}= \left( 1- \frac{R_{S}}{r_{1}} \right)^{1/2}\Delta t,\ \Delta \tau_{2}= \left( 1- \frac{R_{S}}{r_{2}} \right)^{1/2}\Delta t.
\end{equation}

Although the clocks measure different proper times ($\Delta \tau_{1} \neq \Delta \tau_{2}$), the times measured at infinity are the same. Dividing these expressions term by term we obtain

\begin{equation}%66
\frac{\Delta \tau_{1}}{\Delta \tau_{2}}= \frac{\left( 1- \frac{R_{S}}{r_{1}} \right)^{1/2}}{\left( 1- \frac{R_{S}}{r_{2}} \right)^{1/2}}.
\end{equation}

As expected, we see that if $r_{1} \rightarrow \infty$ we recover Eq. (64), since in this case $\Delta \tau_{2} = \Delta \tau$, $\Delta \tau_{1} = \Delta t$, and $r_{2} = r$. We can reach the same conclusions if we assume that $r_{2} > r_{1}$ and take $r_{2} \rightarrow \infty$.

\subsection{Proper time and coordinate time in a circular orbit}

Let us imagine a test particle describing a circular orbit around a massive spherical body. An example of this situation is the orbit described by a planet such as the Earth around the Sun, where the Earth can be considered a test particle since its mass is negligible compared to that of the Sun. We want to relate the proper time measured by a clock traveling with the particle during one complete orbit with the coordinate time interval measured by a distant clock. \\

Under the conditions of the problem, the radius of the orbit will be constant, so that $dr=0$. If we assume that the orbit lies in the equatorial plane, then $\theta=\pi/2$, $d\theta=0$ and $\sin\theta=1$. Imposing these conditions on the Schwarzschild metric we obtain

\begin{equation}%67
c^{2} d\tau^{2} = c^{2}\left( 1-\frac{R_{S}}{r} \right)dt^{2} - r^{2}d\varphi^{2} = c^{2}\left[ 1-\frac{R_{S}}{r}-\frac{r^{2}}{c^{2}}\left( \frac{d\varphi}{dt} \right)^{2}  \right] dt^{2}, 
\end{equation}

where $d\varphi/dt=\omega$ is the angular speed, so that

\begin{equation}%68
d\tau=\left(  1-\frac{R_{S}}{r}-\frac{r^{2}\omega^{2}}{c^{2}} \right)^{1/2} dt.
\end{equation}

Integrating the left-hand side between zero and the orbital proper time $\tau_{o}$, and the right-hand side between zero and the coordinate time (period) $T$, we finally obtain

\begin{equation}%69
\tau_{o}=\left(  1-\frac{R_{S}}{r}-\frac{r^{2}\omega^{2}}{c^{2}} \right)^{1/2} T.
\end{equation}

This expression relates the proper time $\tau_{o}$ indicated by a clock after one complete orbit with the orbital period $T$ measured by a distant observer. Note that if the test particle is stationary, then $\omega=0$, and we recover Eq. (64), as expected.

\subsection{Radius and speed in a circular orbit}

Up to now we have defined a geodesic as the locally straightest path between two points of a manifold. This definition has the advantage of being valid both in Riemannian geometry and in the pseudo-Riemannian geometry of relativistic physics. However, there exists another definition that is not valid in both cases, since it reveals a fundamental difference between these situations: the geodesics of Riemannian geometry are characterized by being the shortest path between two points, whereas timelike geodesics are characterized by the opposite property: they are the longest path between two points in spacetime\footnote{This is a consequence of a fact mentioned earlier: a Riemannian metric is positive definite, whereas a spacetime metric is not.}.\\  

But, as we know, spacetime intervals are directly proportional to proper times, so maximizing this interval is equivalent to maximizing the proper time. Using this idea, and making use of the result from the previous subsection, we can develop a heuristic calculation of the radius and speed of a particle orbiting a massive body. From an educational point of view, the value of this calculation lies in the fact that, although it is heuristic, it leads to the exact result obtained through more complex mathematical procedures. Since the orbit is a geodesic, the orbital radius must maximize the orbital proper time $\tau_{o}$. Applying standard methods of differential calculus, we differentiate Eq. (69) with respect to $r$ and then set the result equal to zero:

\begin{equation}%70
\frac{d\tau_{o}}{dr}=\frac{1}{2}\left(  1-\frac{R_{S}}{r}-\frac{r^{2}\omega^{2}}{c^{2}} \right)^{-1/2}\left( \frac{R_{S}}{r^{2}} - \frac{2r\omega^{2}}{c^{2}} \right)T=0. 
\end{equation}

Therefore,

\begin{equation}%71
\frac{R_{S}}{r^{2}} - \frac{2r\omega^{2}}{c^{2}}=0.
\end{equation}

The solution of this equation for the angular speed is

\begin{equation}%72
\omega= \left( \frac{c^{2}R_{S}}{2r^{3}} \right)^{1/2}= \left( \frac{GM}{r^{3}} \right)^{1/2}.  
\end{equation}

The orbital speed is then

\begin{equation}%73
v=\omega r= \left( \frac{GM}{r} \right)^{1/2}.
\end{equation}

This equality is the same as the expression obtained by applying Newton’s law of universal gravitation\footnote{The magnitude of the centripetal force exerted by a massive object of mass $M$ on a particle of mass $m$ that describes a circular orbit around $M$ is $F=mv^{2}/r$. On the other hand, the magnitude of the gravitational force between $M$ and $m$ is $F=GMm/r^{2}$. Eliminating $F$ between both equations and solving for the orbital velocity we obtain the equation $v=(GM/r)^{1/2}$, which is identical to Eq. (73).}.

\subsection{Null geodesics and photon sphere}

The region of spacetime around a black hole where gravity is so strong that photons are forced to travel in (unstable) circular orbits is called the \textit{photon sphere}. Using the Schwarzschild metric we will find the radius of this orbit. To do so, we proceed in a way analogous to what we did in subsection 9.2, taking $dr=0$ (constant radius) and $\theta=\pi/2$:

\begin{equation} %74
ds^{2} = c^{2}\left( 1 - \frac{R_{S}}{r} \right)dt^{2} - r^{2} d\phi^{2}= \left[ c^{2}\left( 1 - \frac{R_{S}}{r} \right)-r^{2}\omega^{2} \right]dt^{2}.  
\end{equation}

As we know, photons always follow null geodesics in spacetime, for which $ds=0$, Eq. (38), so the previous equation reduces to

\begin{equation}%75
c^{2}\left( 1 - \frac{R_{S}}{r} \right)=r^{2}\omega^{2}=v^{2}.
\end{equation}

Introducing the value of $v$ given by Eq. (73):

\begin{equation}%76
c^{2}\left( 1 - \frac{R_{S}}{r} \right)=\frac{GM}{r}.
\end{equation}

Introducing Eq. (55) and solving for the radius we finally obtain

\begin{equation}%77
r=1,5R_{S} = \frac{3GM}{c^{2}}.
\end{equation}

Evidently, this result is only valid for Schwarzschild black holes. It can be shown that, for a photon to be captured by the photon sphere, an \textit{impact parameter} $b=2,6R_{S}$ is required\footnote{The impact parameter is the perpendicular distance between the trajectory of a projectile (in this case a photon) and the center of a force field (such as the gravitational field generated by the black hole).}. For $b>2,6R_{S}$ the photon can escape the gravity of the black hole, and for $b<2,6R_{S}$ it is absorbed by the black hole [19].

\section{The Schwarzschild metric and embedding diagrams}

Visualizing a four-dimensional curved spacetime such as the one proposed by relativistic physics is a task that not even Einstein himself could accomplish. However, physicists have devised an ingenious way to circumvent this difficulty by embedding a two-dimensional version of the Schwarzschild metric described by the spherical coordinates $(r,\varphi)$ into a three-dimensional space described by the cylindrical coordinates $(z,r,\varphi)$. In other words, the basic idea is to embed a two-dimensional space into a three-dimensional one [20,21]. What do we gain from this?\\ 

Let us imagine a two-dimensional being living on the surface of a sphere from which it cannot escape. Evidently, it will not be able to visualize the curvature or the properties of its two-dimensional world. But a three-dimensional being like us, who can see the sphere from the outside, embedded in a three-dimensional space, will notice how the surface curves into a third dimension. We can do something analogous with the Schwarzschild spacetime geometry by means of the so-called \textit{embedding diagram} [1,17].\\  

To construct the diagram from the Schwarzschild metric, we must recall that the metric is stationary and static. This means that the Schwarzschild geometry looks exactly the same at any instant, which allows us to take $t=constant$, so that $dt=0$. Moreover, since the Schwarzschild geometry is spherically symmetric, all planes passing through the center are equivalent, and for simplicity we can consider the equatorial plane by taking $\theta=90^{o}$, so that $\sin \theta=1$ and $d\theta=0$. Imposing these conditions and considering the signature $\lbrace - + + + \rbrace$, the four-dimensional solution, Eq. (54), reduces to a two-dimensional metric defined by the coordinates $(r,\varphi)$:

\begin{equation}%78
ds^{2}=\left( 1-\frac{R_{S}}{r} \right)^{-1} dr^{2}+r^{2}d\varphi^{2}. 
\end{equation}

This expression describes the Schwarzschild geometry in the equatorial plane (Fig. 11) as a function of the coordinate $r$. Thus, each value of $r$ and each rotation of $\varphi$ by $2\pi$ define a circumference centered on the $z$ axis. This means that cylindrical symmetry exists, which reveals that we can embed the two-dimensional space $(r,\varphi)$ into a three-dimensional one described by the cylindrical coordinates $(z,r,\varphi)$. The metric of a three-dimensional space in cylindrical coordinates is a simple extension along the $z$ axis of the two-dimensional Euclidean metric in polar coordinates, Eq. (7), and is given by the expression

\begin{equation}%79
ds^{2}=dz^{2}+dr^{2}+r^{2}d\varphi^{2}.
\end{equation}

This equality can be rewritten as

\begin{equation}%80
ds^{2}=\left( \frac{dz}{dr} \right)^{2} dr^{2}+dr^{2}+r^{2}d\varphi^{2}= \left[ \left( \frac{dz}{dr} \right)^{2} +1 \right]dr^{2} +r^{2}d\varphi^{2}.
\end{equation}

If Eqs. (78) and (80) represent the same geometry, then it must hold that 

\begin{equation}%81
\left( \frac{dz}{dr} \right)^{2} +1=\left( 1-\frac{R_{S}}{r} \right)^{-1}.
\end{equation}

Therefore: 

\begin{equation}%82
\frac{dz}{dr}=\pm \left( \frac{r}{R_{S}} -1 \right)^{-1/2}. 
\end{equation}

Integrating this expression between the corresponding limits we obtain 

\begin{equation} %83
z = \pm \int_{0}^{r} \dfrac{dr}{\sqrt{\frac{r}{R_{S}} - 1}} = \pm 2\sqrt{R_{S}(r-R_{S})}.
\end{equation}

The equation 

\begin{equation}%84
z= \pm 2\sqrt{R_{S}(r-R_{S})}, 
\end{equation}

corresponds to a parabola that intersects the $r$ axis at $R_{S}$, so that this axis is expressed in units of the Schwarzschild radius, as illustrated in Fig. 12. To the right of the $z$ axis in the figure appears the parabola in blue, together with the same parabola rotated $180^{o}$, which appears to the left of the $z$ axis. The upper region of the parabola ($z^{+}$) and the lower region ($z^{-}$) are also shown.\\  

\begin{figure}[h]
  \centering
    \includegraphics[width=0.55\textwidth]{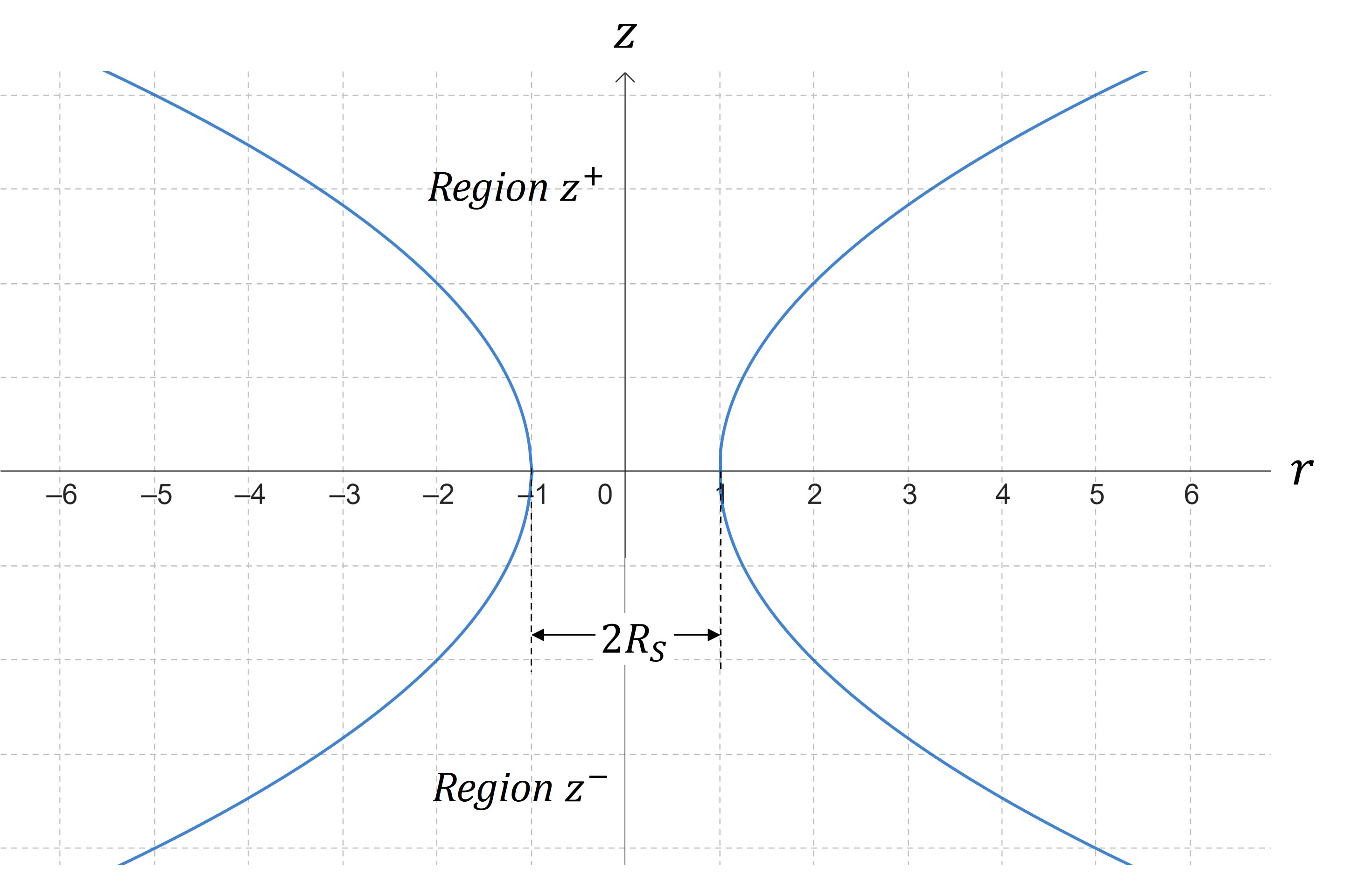}
  \caption{Parabola whose continuous rotation around the $z$ axis generates the two-dimensional Schwarzschild geometry.}
\end{figure} 

The continuous rotation of this parabola around the $z$ axis generates a curved two-dimensional surface called the \textit{Flamm paraboloid}, which allows us to visualize the Schwarzschild geometry, illustrated in Fig. 13 [1,17]. The region $z^{+}$ represents the spacetime of ordinary objects that compose the known universe. In particular, we know that for $r>R_{S}$, the region $z^{+}$ corresponds to the spacetime exterior to a Schwarzschild black hole. The diagram does not show the geometry interior to $R_{S}$, because for $r<R_{S}$ the value of $z$ in Eq. (84) is complex.\\  

We see that Eq. (84) admits a negative solution corresponding to the region $z^{-}$, which also appears illustrated in Fig. 13. By joining the positive solution $z^{+}$ and the negative solution $z^{-}$ we obtain a \textit{wormhole}, also called an \textit{Einstein–Flamm–Rosen bridge} in honor of the physicists who discovered it (Fig. 13). This object possesses a \textit{throat} (physically non-traversable) of radius $R_{S}$ that separates two universes, an upper one and a lower one. Due to space limitations we cannot analyze the properties of wormholes in more detail. The interested reader can find excellent nontechnical discussions in [21,22].\\

\begin{figure}[h]
  \centering
    \includegraphics[width=0.4\textwidth]{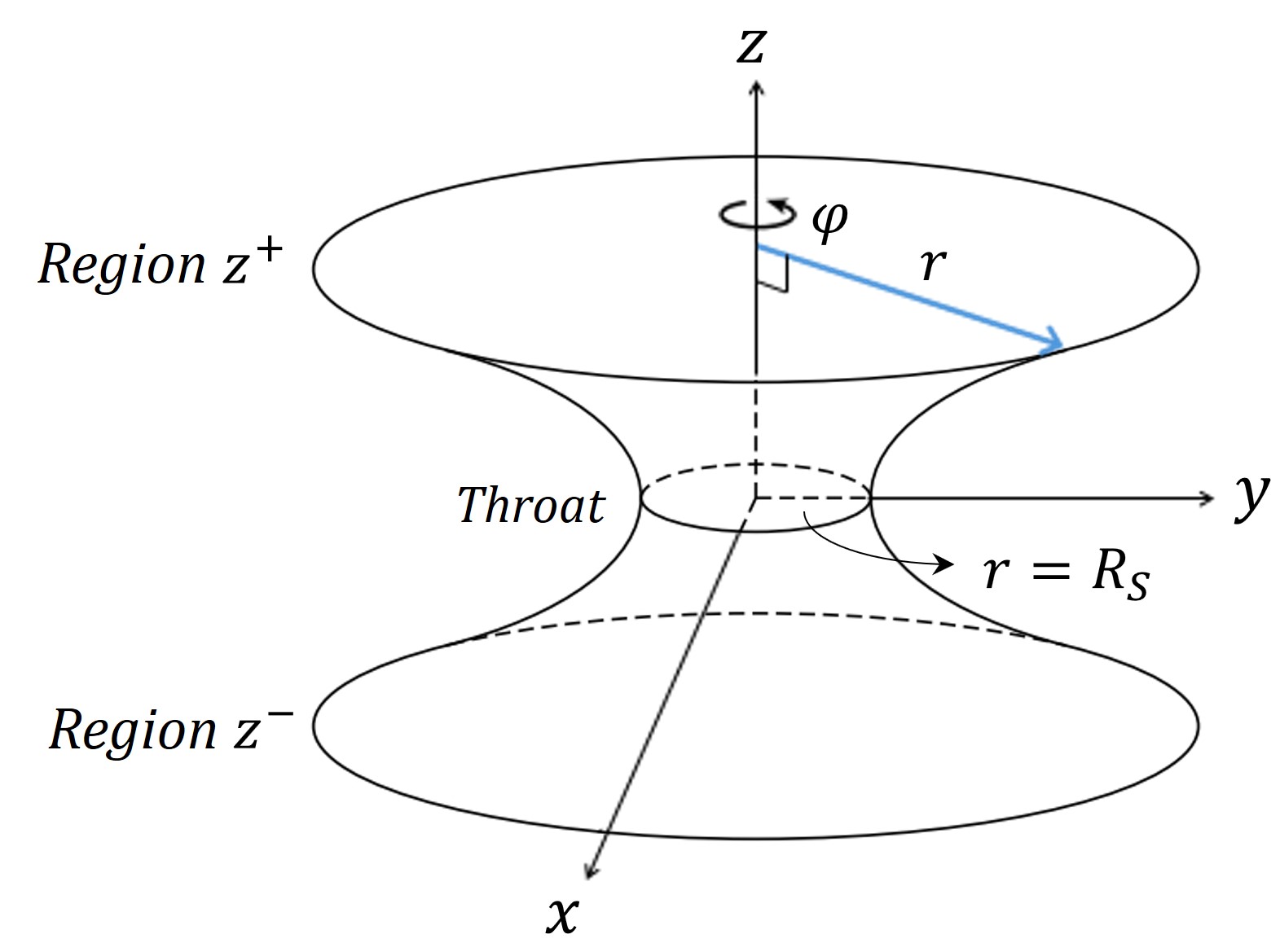}
  \caption{The Flamm paraboloid, which represents Schwarzschild spacetime.}
\end{figure} 

It is important to keep in mind that only the two-dimensional surface of the paraboloid has physical meaning as part of the Schwarzschild geometry. No trajectory that leaves the surface is allowed. The three-dimensional Euclidean space in which the surface is embedded is only a tool to visually represent curved spacetime [1,17]. The Flamm paraboloid shows the entire Schwarzschild spacetime for $r>R_{S}$, and if the mass is contained in the region $r<R_{S}$, then it shows the spacetime outside the horizon of a Schwarzschild black hole whose mass is concentrated at $r=0$, where a physical singularity exists. Due to the Jebsen-Birkhoﬀ theorem, mentioned in section 8, this external space-time is identical to that of a Schwarzschild black hole, regardless of whether the collapsing mass has already concentrated at r = 0, where a physical singularity exists.\\

If we want to observe the spacetime exterior to a spherical celestial body of radius $R$, such as the Sun, we must consider the region of the paraboloid for which $r>R$. As in the case of the Sun, $R_{\odot} \cong 7\times 10^{5}km$, and $R_{S}\cong 3km$ then

\begin{equation}%85
\frac{R_{\odot}}{R_{S}} \cong 2,3 \times 10^{5},
\end{equation}

so that for $r>2,3 \times 10^{5}$ in units of the Schwarzschild radius, the diagram represents the spacetime exterior to the Sun. The Flamm paraboloid contains valuable geometrical and physical information about Schwarzschild spacetime, and it can be extracted directly from Fig. 13 using spatial intuition, without the need to perform complicated calculations. For example, we see that spacetime is asymptotically flat, which means that as we move away from $r=R_{S}$, spacetime becomes increasingly flat. We can obtain a simple intuitive measure of the curvature in the radial direction of Schwarzschild spacetime by considering the slope of the paraboloid given by Eq. (82), where we are only interested in the positive solution: 

\begin{equation}%86
\frac{dz}{dr}= \left( \frac{r}{R_{S}} -1 \right)^{-1/2}= \frac{1}{\sqrt{\frac{r}{R_{S}}-1}}. 
\end{equation}

We see, for example, that for $r/R_{S}\rightarrow  \infty$, $dz/dr \rightarrow0$. That is, for $r\gg R_{S}$ the slope is zero and the surface of the paraboloid is flat. If we want to calculate the curvature in the region just outside the solar surface, we take $r/R_{S}=R_{\odot}/R_{S} \cong 2,3\times 10^{5}$ in Eq. (86), obtaining $dz/dr \cong 2\times 10^{-3}$. This is an extremely small slope and shows that the spacetime curvature around the Sun is very small, reinforcing a conclusion obtained earlier, which we can now visualize: Newton’s law of gravitation is a good approximation for describing the solar system.\\

\begin{figure}[h]
  \centering
    \includegraphics[width=0.4\textwidth]{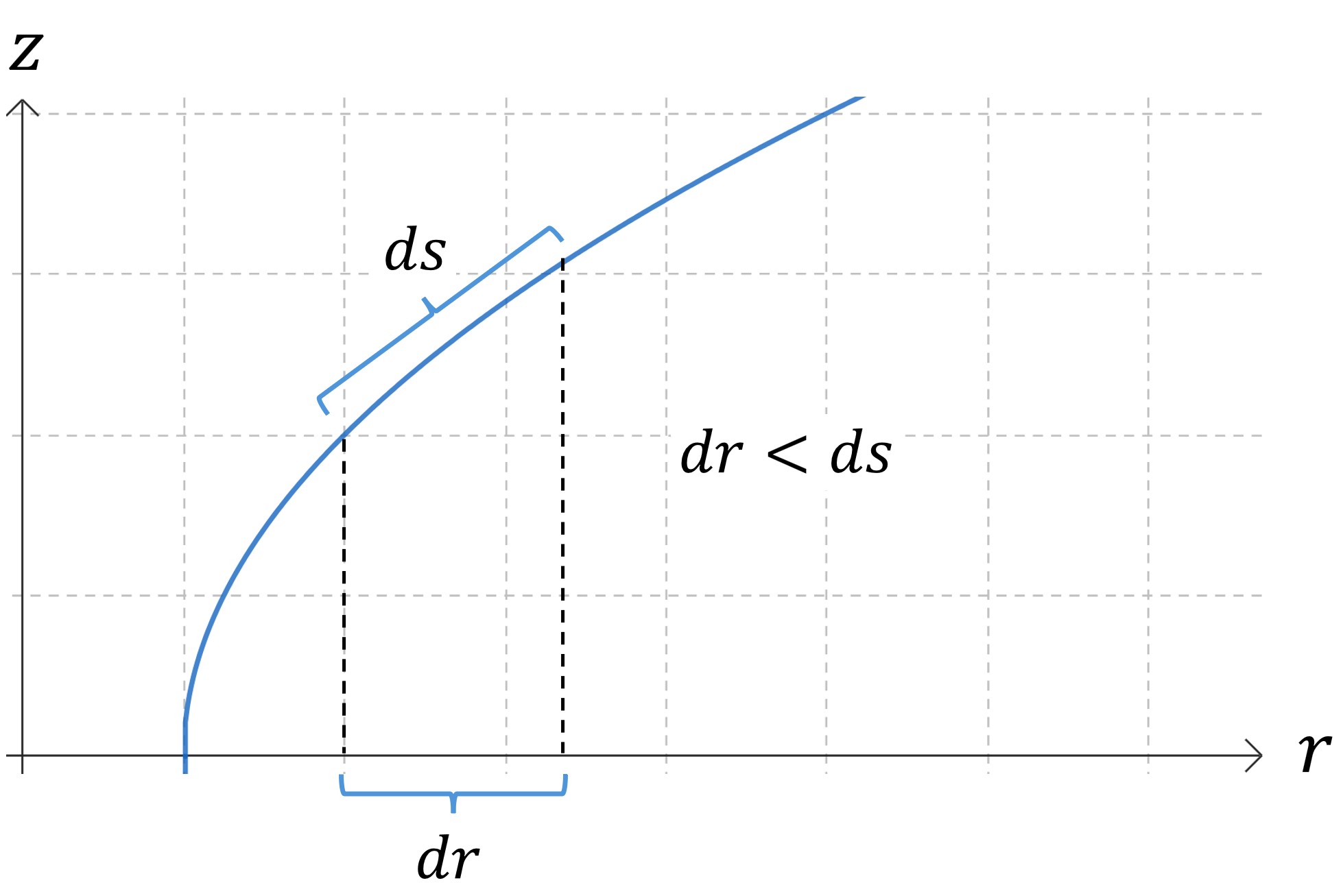}
  \caption{Comparison between the distance measured along the $r$ axis and along the Flamm paraboloid.}
\end{figure} 

Another interesting aspect of Schwarzschild geometry revealed by the Flamm paraboloid is that radial distances are stretched in the reference frame of an observer approaching $r=R_{S}$. As shown in Fig. 14, the distance $ds$ between any two points measured on the paraboloid is greater than the corresponding distance $dr$ measured along the radial coordinate $r$. We can reach this conclusion from Eq. (78), where it is observed that if we move only in the radial direction, then $d\varphi=0$, so that 

\begin{equation}%87
ds=\frac{dr}{\sqrt{1-\frac{R_{S}}{r}}}.
\end{equation}

We see that, for $r>R_{S}$, $ds>dr$. This result visually reinforces another idea mentioned earlier: near the Schwarzschild radius the coordinate $r$ does not have a direct physical meaning, and therefore cannot be interpreted as a physical distance, except when $r \rightarrow \infty$, since in that case Eq. (87) shows that $ds=dr$.\\

As the reader may have noticed, the shape of the Flamm paraboloid resembles the well-known \textit{rubber sheet analogy}, a pedagogical resource widely used to represent the curvature of spacetime produced by gravitating bodies. However, the similarity between the two representations is superficial. The most obvious difference is that only the Flamm paraboloid is mathematically rigorous, but there are other differences. One of them in particular deserves special attention: the paraboloid represents Schwarzschild geometry at constant time, that is, it is a kind of snapshot of spacetime, which implies that no objects can move along the paraboloid; however, in the rubber sheet analogy teachers usually show objects rolling freely across the elastic surface. We leave it to the reader to discover other similarities and differences between the Flamm paraboloid and the rubber sheet analogy.

\section{Final comments}

Research in GR is experiencing an exciting era, full of challenges and new discoveries, creating a favorable scenario for explaining and disseminating Einstein’s theory among undergraduate students. However, the great mathematical complexity of GR often makes this task difficult, relegating the study of this subject to the graduate level. But we have seen that this difficulty can be easily overcome if we take advantage of the educational potential of the Schwarzschild solution, which, besides being the gateway to GR, has a great diversity of physical and astronomical applications whose understanding is within reach of any undergraduate science or engineering student who is willing to make a small effort. The introductory analysis of the Schwarzschild solution that we have presented can be especially useful as a prelude to the study of GR in graduate courses, where it is highly advisable to develop the students’ physical and geometrical intuition before they fully immerse themselves in the sophisticated mathematics of Einstein’s theory. \\

A little more than a century has passed since Schwarzschild found his exact solution, and the knowledge accumulated during this time is formidable. This has forced us to be selective, and it has not been easy to determine which topics to include and which to exclude. A current topic that has only been briefly presented is the physics of black holes, whose mathematical foundations lie in the Schwarzschild solution. Likewise, there are other applications of the Schwarzschild metric that we have not analyzed, such as the study of light cones in the surroundings of a spherical mass distribution, a topic that provides a beautiful geometrical perspective of spacetime that complements the information provided by embedding diagrams. The reader who wishes to explore this and other topics not addressed can find in these pages the foundations to continue with broader and deeper studies on gravitation and GR. In any case, it is to be hoped that the topics discussed will become stimulating material for analysis and discussion, both for students interested in GR and for instructors who teach courses related to this important topic, especially at the undergraduate level.

\subsection*{Appendix A. Tensors and coordinate transformations}

One of the fundamental ideas of Riemannian geometry is that the geometric properties of a manifold must not depend on the coordinate system used to describe it. In particular, the squared line element, $ds^{2}$, must remain invariant under coordinate transformations. This requirement determines how the metric tensor transforms between different coordinate systems. Below we analyze a simple example that illustrates the systematic procedure used in Riemannian geometry to perform a coordinate transformation. \\

Let us consider a flat two-dimensional space described in orthogonal Cartesian coordinates. The metric is written as

\begin{equation} \tag{A1}
ds^{2}= dx^{2}+dy^{2}.
\end{equation}

More generally, the squared line element can be expressed as

\begin{equation} \tag{A2}
ds^{2}=g_{ij}dx^{i}dx^{j},
\end{equation}

where \(g_{ij}\) are the components of the metric tensor. In Cartesian coordinates we have

\[
g_{11}=g_{22}=1,
\qquad
g_{12}=g_{21}=0.
\]

Suppose now that we introduce a new coordinate system \(x'^k\). The coordinate differentials transform according to the chain rule,

\begin{equation} \tag{A3}
dx^{i}=
\frac{\partial x^{i}}{\partial x'^{k}}
dx'^{k}.
\end{equation}

Substituting Eq.~(A3) into Eq.~(A2), we obtain

\begin{equation} \tag{A4}
ds^{2}
=
g_{ij}
\frac{\partial x^{i}}{\partial x'^{k}}
\frac{\partial x^{j}}{\partial x'^{l}}
dx'^{k}dx'^{l}.
\end{equation}

Defining

\begin{equation} \tag{A5}
g'_{kl}
=
\frac{\partial x^{i}}{\partial x'^{k}}
\frac{\partial x^{j}}{\partial x'^{l}}
g_{ij},
\end{equation}

Eq.~(A4) becomes

\begin{equation} \tag{A6}
ds^{2}=g'_{kl}dx'^{k}dx'^{l}.
\end{equation}

Equation~(A6) shows that the squared line element has exactly the same form in the new coordinate system. Therefore, the quantity \(ds^{2}\) is invariant under coordinate transformations, while the metric components transform according to Eq.~(A5). This transformation law implies that the metric tensor is a second-order covariant tensor. In Eq.~(A5), \(k,l\) are called \textit{free indices}, while \(i,j\) are known as \textit{dummy indices}, and the Einstein summation convention is applied to them. \\

To illustrate these ideas, let us consider the transformation from Cartesian coordinates \((x,y)\) to polar coordinates \((r,\varphi)\),

\begin{equation} \tag{A7}
x=r\cos\varphi,
\qquad
y=r\sin\varphi,
\end{equation}

where

\[
x^{1}=x,
\qquad
x^{2}=y,
\qquad
x'^{1}=r,
\qquad
x'^{2}=\varphi.
\]

From Eq.~(A7) we obtain

\begin{equation} \tag{A8}
\frac{\partial x}{\partial r}=\cos\varphi,
\qquad
\frac{\partial y}{\partial r}=\sin\varphi,
\qquad
\frac{\partial x}{\partial \varphi}=-r\sin\varphi,
\qquad
\frac{\partial y}{\partial \varphi}=r\cos\varphi.
\end{equation}

Applying Eq.~(A5), and considering the Einstein summation convention, we obtain

\begin{equation} \tag{A9}
g'_{11}
=
\left(
\frac{\partial x}{\partial r}
\right)^{2}
+
\left(
\frac{\partial y}{\partial r}
\right)^{2}
=
\cos^{2}\varphi+\sin^{2}\varphi
=
1,
\end{equation}

and

\begin{equation} \tag{A10}
g'_{22}
=
\left(
\frac{\partial x}{\partial \varphi}
\right)^{2}
+
\left(
\frac{\partial y}{\partial \varphi}
\right)^{2}
=
r^{2}
(\sin^{2}\varphi+\cos^{2}\varphi)
=
r^{2}.
\end{equation}

We leave as an exercise for the reader to show that

\[
g'_{12}=g'_{21}=0.
\]

Substituting these results into Eq.~(A6), and considering that \(x'^{1}=r\) and \(x'^{2}=\varphi\), we finally obtain

\begin{equation} \tag{A11}
ds^{2}=dr^{2}+r^{2}d\varphi^{2}.
\end{equation}

Thus, the Euclidean plane can be described either by Cartesian coordinates or by polar coordinates. The metric components are different in each coordinate system, but the geometric interval \(ds^{2}\) remains invariant. \\

Although the procedure used to go from the metric in Cartesian coordinates to the metric in polar coordinates may seem very sophisticated for a problem that could be solved using elementary geometrical methods, in GR it is common to perform coordinate transformations whose complexity requires a systematic and rigorous procedure such as the one described here. Tensor calculus, of which we have seen a simple application, is the mathematical language par excellence in which GR is formulated.

\subsection*{Appendix B. Solution of Einstein’s equation for a flat spacetime}

Let us return to Eq. (41). This equation is written explicitly in the form:

\begin{equation} \tag{B1}
R_{\mu\nu}-\frac{1}{2}Rg_{\mu\nu}=0.
\end{equation}

We want to verify that the Minkowski metric tensor, Eq. (35), is a solution of this equation. To do so we must recall that the Ricci tensor is defined by the equality

\begin{equation} \tag{B2}
R_{\mu\nu}= R_{\mu\nu\lambda}^{\lambda},
\end{equation}

where $R_{\mu\nu\lambda}^{\lambda}$ is the contracted Riemann tensor, which in general is defined as

\begin{equation} \tag{B3}
R_{\sigma\mu\nu}^{\rho} = \dfrac{\partial \Gamma _{\sigma\nu}^{\rho}}{\partial x^{\mu}}-\dfrac{\partial \Gamma_{\sigma\mu}^{\rho}}{\partial x^{\nu}}+ \Gamma _{\sigma\nu}^{\lambda} \Gamma _{\lambda\mu}^{\rho} - \Gamma _{\sigma\mu}^{\lambda} \Gamma _{\lambda\nu}^{\rho}.
\end{equation}

The quantities $\Gamma _{\mu\nu}^{\sigma}$ are called \textit{Christoﬀel symbol of the second kind}, and they are defined from the partial derivatives of the metric components:

\begin{equation} \tag{B4}
\Gamma _{\mu\nu}^{\sigma}= \frac{1}{2}g^{\sigma\rho} \left\lbrace  \dfrac{\partial g_{\rho\nu}}{\partial x^{\mu}} + \dfrac{\partial g_{\mu\rho}}{\partial x^{\nu}} - \dfrac{\partial g_{\mu\nu}}{\partial x^{\rho}} \right\rbrace. 
\end{equation}

Since the components of the Minkowski metric are constant on the diagonal and vanish in the remaining cases, the derivatives are zero and the connection coefficients also vanish

\begin{equation} \tag{B5}
\Gamma _{\mu\nu}^{\sigma}= 0.
\end{equation}

This implies that the Riemann tensor vanishes. Then, from Eq. (B3), we have that the Ricci tensor reduces to

\begin{equation} \tag{B6}
R_{\mu\nu}= 0.
\end{equation}

On the other hand, the Ricci scalar is defined as

\begin{equation} \tag{B7}
R = g^{\mu\nu}R_{\mu\nu},
\end{equation}

so that, from Eq. (B6), we obtain

\begin{equation} \tag{B8}
R = 0.
\end{equation}

Substituting Eqs. (B6) and (B8) into Eq. (B1) we see that the equality is satisfied, which shows that the Minkowski metric is a solution of this equation.

\subsection*{Appendix C. The Schwarzschild vacuum solution}

As we know, in order to find the Schwarzschild vacuum solution we must begin by setting the energy–momentum tensor to zero in Einstein’s equation:

\begin{equation} \tag{C1}
R_{\mu\nu}-\frac{1}{2}Rg_{\mu\nu}=0.
\end{equation}

Let us multiply both sides of this equality by the contravariant form of the metric tensor:

\begin{equation} \tag{C2}
g^{\mu\nu} \left( R_{\mu\nu}-\frac{1}{2}Rg_{\mu\nu} \right) = g^{\mu\nu}R_{\mu\nu}-\frac{1}{2}R(g^{\mu\nu}g_{\mu\nu})=0,
\end{equation}

where, by contraction of the indices $\mu$ and $\nu$, we obtain

\begin{equation} \tag{C3}
g^{\mu\nu}R_{\mu\nu}= R_{\nu}^{\nu}=R_{0}^{0}+R_{1}^{1}+R_{2}^{2}+R_{3}^{3}\equiv R.
\end{equation}

By contraction of the same indices, we obtain

\begin{equation} \tag{C4}
g^{\mu\nu}g_{\mu\nu}=\delta_{\nu}^{\nu}=\delta_{0}^{0}+\delta_{1}^{1}+\delta_{2}^{2}+\delta_{3}^{3} = 4,
\end{equation}

where a quantity called the \textit{Kronecker delta} has been introduced, defined as

\begin{equation} \tag{C5}
\delta^{\mu}_{\nu}=  
\begin{cases}
0 & \text{if $\mu \neq \nu$} \\
1 & \text{if $\mu =\nu$}
\end{cases}.
\end{equation}

Substituting Eqs. (C3) and (C4) into Eq. (C1) we obtain $R-(4R)/2=-R=0$, which implies that $R=0$, so that Eq. (C1) finally reduces to

\begin{equation} \tag{C6}
R_{\mu\nu}=0.
\end{equation}

Both the Ricci tensor and the Ricci scalar must vanish for a vacuum solution. However, this condition alone does not completely determine the geometry of spacetime; it is necessary to explicitly solve Eq. (C6) in order to obtain the corresponding metric, finding the values of $R_{\mu\nu}$ for all pairs of indices $\mu$ and $\nu$.

\section*{Data Availability}
The entire dataset that supports the results of this study was published in the article itself.

\section*{References}

[1]	A.M. Steane, Relativity Made Relatively Easy, Oxford University Press, Oxford, 2012.

\vspace{2mm}

[2]	J. Pinochet, General relativity in a nutshell II, Phys. Scr. 98 (2023) 126104. https://doi.org/10.1088/1402-4896/ad0c15.

\vspace{2mm}

[3]	J.M. Sánchez-Ron, Albert Einstein: Su vida, su obra y su mundo, Crítica, Buenos Aires, 2016.

\vspace{2mm}

[4]	R. Lambourne, Relativity, Gravitation and Cosmology, Cambridge University Press, 2010.

\vspace{2mm}

[5]	J. Pinochet, General relativity in a nutshell I, Phys. Scr. 98 (2023) 126103. https://doi.org/10.1088/1402-4896/ad0c34.

\vspace{2mm}

[6]	B. Schutz, A First Course in General Relativity, Cambridge University Press, Cambridge, 2009.

\vspace{2mm}

[7]	J.A. Wheeler, A journey into gravity and spacetime, W. H. Freeman and Company, New York, 1990.

\vspace{2mm}

[8]	J.B. Hartle, Gravity: An introduction to Einstein’s General Relativity, Addison Wesley, San Francis-co, 2003.

\vspace{2mm}

[9]	V. Faraoni, Special Relativity, Springer, New York, 2013.

\vspace{2mm}

[10] J. Pinochet, Classical Tests of General Relativity Part I: Looking to the Past to Understand the Pre-sent, Physics Education 55 (2020) 65016.

\vspace{2mm}

[11] S. Hacyan, Relatividad para estudiantes de física, Fondo de Cultura Económica, México, D.F., 2013.

\vspace{2mm}

[12] M.P. Hobson, G.P. Efstathiou, A.N. Lasenby, General Relativity. An Introduction for Physicists, Cambridge University Press, Cambridge, 2006.

\vspace{2mm}

[13] H. Stephani, D. Kramer, M. MacCallum, C. Hoenselaers, E. Herlt, Exact Solutions of Einstein’s Field Equations, Cambridge University Press, Cambridge, 2003.

\vspace{2mm}

[14] V.P. Frolov, A. Zelnikov, Introduction to Black Hole Physics, Oxford University Press, Oxford, 2011.

\vspace{2mm}

[15] T.L. Chow, Gravity, Black Holes, and the Very Early Universe, Springer, New York, 2008.

\vspace{2mm}

[16] K. Schwarzschild, On the Gravitational Field of a Mass Point according to Einstein’s Theory, translated by S. Antoci and A. Loinger, arXiv:physics/9905030 (1999).

\vspace{2mm}

[17] W. Rindler, Relativity: Special, General and Cosmological, Oxford University Press, New York, 2006.

\vspace{2mm}

[18] C. Bambi, Introduction to General Relativity A Course for Undergraduate Students of Physics, Springer, Singapore, 2018.

\vspace{2mm}

[19] H. Cohn, Black hole physics illustrated in photon orbits, Am. J. Phys. 45 (1977) 239–241.

\vspace{2mm}

[20] K.S. Thorne, Agujeros negros y tiempo curvo: El escandaloso legado de Einstein, Crítica, Barcelo-na, 2000.

\vspace{2mm}

[21] J.P. Luminet, Agujeros negros, Alianza, Madrid, 1991.

\vspace{2mm}

[22] M.S. Morris, K.S. Thorne, Wormholes in spacetime and their use for interstellar travel: A tool for teaching general relativity, Am. J. Phys. 56 (1988) 395–412.

\end{document}